\documentclass[aps,prl,reprint,superscriptaddress,amssymb,nobibnotes, 11pt, twocolumn]{revtex4-2}

\pdfoutput=1
\usepackage{amsmath, amsfonts}    
\usepackage{graphicx}   
\usepackage{verbatim}   
\usepackage{color}      
\usepackage{subfigure}  
\usepackage{hyperref}   
\usepackage[font=small]{caption}
\usepackage[english]{babel}

\usepackage{natbib} 

\usepackage[utf8]{inputenc} 
\DeclareUnicodeCharacter{2009}{\,} 

\newcommand{\ket}[1]{|#1\rangle}
\newcommand{\braket}[2]{\langle #1|#2\rangle}

\makeatletter
\def\maketitle{
\@author@finish
\title@column\titleblock@produce
\suppressfloats[t]}
\makeatother

\begin{document}

\title{Qubit teleportation between a memory-compatible photonic time-bin qubit and a solid-state quantum network node}

\author{Mariagrazia Iuliano}\thanks{These authors contributed equally to this work}\affiliation{QuTech \& Kavli Institute of Nanoscience Delft, Delft University of Technology, Delft, The Netherlands} \author{Marie-Christine Slater}\thanks{These authors contributed equally to this work}\affiliation{QuTech \& Kavli Institute of Nanoscience Delft, Delft University of Technology, Delft, The Netherlands} \author{Arian J. Stolk}\affiliation{QuTech \& Kavli Institute of Nanoscience Delft, Delft University of Technology, Delft, The Netherlands} \author{Matthew J. Weaver}\affiliation{QuTech \& Kavli Institute of Nanoscience Delft, Delft University of Technology, Delft, The Netherlands} \author{Tanmoy Chakraborty}\affiliation{QuTech \& Kavli Institute of Nanoscience Delft, Delft University of Technology, Delft, The Netherlands} \author{Elsie Loukiantchenko}\affiliation{QuTech \& Kavli Institute of Nanoscience Delft, Delft University of Technology, Delft, The Netherlands} \author{Gustavo Castro do Amaral}\affiliation{QuTech \& Kavli Institute of Nanoscience Delft, Delft University of Technology, Delft, The Netherlands} \author{Nir Alfasi}\affiliation{QuTech \& Kavli Institute of Nanoscience Delft, Delft University of Technology, Delft, The Netherlands} \author{Mariya O. Sholkina}\affiliation{QuTech \& Kavli Institute of Nanoscience Delft, Delft University of Technology, Delft, The Netherlands} \author{Wolfgang Tittel}\affiliation{Department of Applied Physics, University of Geneva, Geneva, Switzerland}\affiliation{Constructor University Bremen GmbH, Bremen, Germany}
\author{Ronald Hanson}\email{r.hanson@tudelft.nl} \affiliation{QuTech \& Kavli Institute of Nanoscience Delft, Delft University of Technology, Delft, The Netherlands}

\begin{abstract}
We report on a quantum interface linking a diamond NV center quantum network node and 795nm photonic time-bin qubits compatible with Thulium and Rubidium quantum memories. The interface makes use of two-stage low-noise quantum frequency conversion and waveform shaping to match temporal and spectral photon profiles. Two-photon quantum interference shows high indistinguishability of (89.5$\pm$1.9)\% between converted 795nm photons and the native NV center photons. We use the interface to demonstrate quantum teleportation including real-time feedforward from an unbiased set of 795nm photonic qubit input states to the NV center spin qubit, achieving a teleportation fidelity of (75.5$\pm$1.0)\%. This proof-of-concept experiment shows the feasibility of interconnecting different quantum network hardware.
\end{abstract}
\maketitle
\section{Introduction}
The future quantum internet will leverage the principles of quantum mechanics for ultra-secure communication, enhanced sensing, and distributed quantum computing \cite{kimble_quantum_2008, wehner_quantum_2018}. Progress in the past decade has led to pioneering experiments on different components of such a network \cite{togan_quantum_2010, lodahl_quantum-dot_2017, bock_high-fidelity_2018, ruf_quantum_2021, stas_robust_2022}. For instance, entanglement generation between separated quantum memory systems based on atomic ensembles has recently been reported \cite{lago-rivera_long_2023, liu_multinode_2023} and the first multi-node network of rudimentary quantum processors has been realized inside the lab \cite{pompili_realization_2021, hermans_qubit_2022}. As different hardware platforms may be optimized for different network tasks, realizing interfaces that enable quantum information transfer between heterogeneous devices is a key challenge.

Here, we report on a proof-of-concept demonstration of a quantum interface between a diamond NV center qubit \cite{doherty_nitrogen-vacancy_2013, childress_diamond_2013} and photonic time-bin qubits at 795nm that are compatible with Thulium-based solid-state memories \cite{thiel_tm3y3ga5o12_2014, sinclair_proposal_2016, askarani_long-lived_2021} and Rubidium-based atomic gas memories \cite{cho_highly_2016, zhao_long-lived_2009, rosenfeld_remote_2007, heller_cold-atom_2020}. Such an interface conceptually corresponds to future quantum Internet scenarios such as connecting remote qubit processors via a repeater chain \cite{azuma_quantum_2023} or realizing remote state preparation on a quantum computing server from a photonic client \cite{drmota_verifiable_2023}. We validate the quantum nature of the interface by performing quantum teleportation \cite{bennett_teleporting_1993, bouwmeester_experimental_1997} of 795nm time-bin qubits into the NV center spin qubit with state fidelity beating the classical bound.
\begin{figure}[htp]
    \includegraphics[width=1.0\columnwidth]{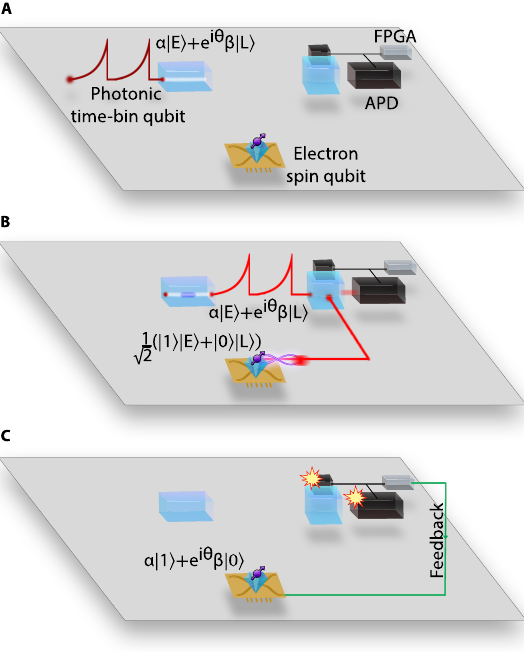}
    \caption{\textbf{Overview of the quantum interface between 795nm photonic time-bin qubits and an NV center processing node.} The interface consists of a low-noise two-step frequency conversion module including a frequency stabilization module, an interference station containing a balanced beam splitter with output ports connected to avalanche photodiodes (APDs), and an FPGA for real-time feedback. The interface can be visualized in three steps: a) a 795nm time-bin qubit with a temporal shape matching the spontaneous emission profile of the NV center is sent to the input of the interface. The NV spin qubit is prepared in a balanced superposition state. b) The 795~nm photonic qubit is converted to 637~nm, while the NV center generates a 637~nm photonic time-bin qubit entangled with the spin qubit. The generation of the 795 nm photonic qubit is timed to ensure maximum overlap at the beam splitter with the NV center photonic qubit. (c) Upon detecting one photon in each time bin, feedback of the correct phase flip to the NV spin qubit of the NV completes the state teleportation. For the experiments reported in this paper, we employed the NV qubit setup, called ``Alice", described in Refs.~\cite{pompili_realization_2021, hermans_qubit_2022}, that includes a micro-controller unit (MCU) and a fast waveform generator (AWG).
    }
    \label{fig:fig1}
\end{figure}
\subsection{A quantum interface between heterogeneous devices}
A major challenge for linking heterogeneous quantum network hardware is the matching of their corresponding photonic qubits. Many leading hardware platforms for quantum memories and quantum network nodes are based on atom-like systems \cite{heshami_quantum_2016, lei_quantum_2023}. The properties of the photonic interface of these platforms, such as temporal profile and wavelength of emitted photons, are therefore largely determined by the atomic properties and vary significantly among the different platforms. Our approach to bridging these differences is depicted in the schematic of our interface in Fig. \ref{fig:fig1}.

The interface converts the input 795nm photonic time-bin qubit to match the properties of the NV center photon. In parallel, entanglement is generated between the spin state of the NV center and the temporal mode of a single emitted photon. Then, the converted 795nm photon and the NV photon are interfered on a beam-splitter. Subsequent detection of the photons in different time bins constitutes a Bell state measurement that teleports the original 795nm time-bin qubit state to the NV spin qubit. Real-time feedforward of the Bell-state measurement outcome and application of the corresponding correction gate on the NV spin qubit completes the action of the interface.

For this interface to function with high fidelity, it is crucial that the converted 795nm photons are indistinguishable from the NV center photons. In particular, the 795nm photons need to match the NV photons' 637nm wavelength, polarization and exponential temporal profile set by NV's 12~ns optical lifetime. In this proof-of-concept work, we create photonic time-bin qubits at 795nm from weak coherent states by using an intensity modulator and a phase modulator. We calibrate the intensity modulator to mimic the NV photon's temporal profile within a 30ns time window. The photonic states obtained through this method are compatible with the storage and retrieval from Thulium-doped solid-state quantum memories \cite{afzelius_multimode_2009} as well as Rubidium-gas-based quantum memories \cite{covey_quantum_2023, lipka_massively-multiplexed_2021}. These platforms are capable of multiplexing by storage and retrieval of multiple photonic modes in different degrees of freedom \cite{bussieres_prospective_2013, lei_quantum_2023}, and therefore have attracted interest for quantum repeater applications \cite{sangouard_quantum_2011}.

\begin{figure*}[htp]
  \includegraphics[width=1.0\linewidth]{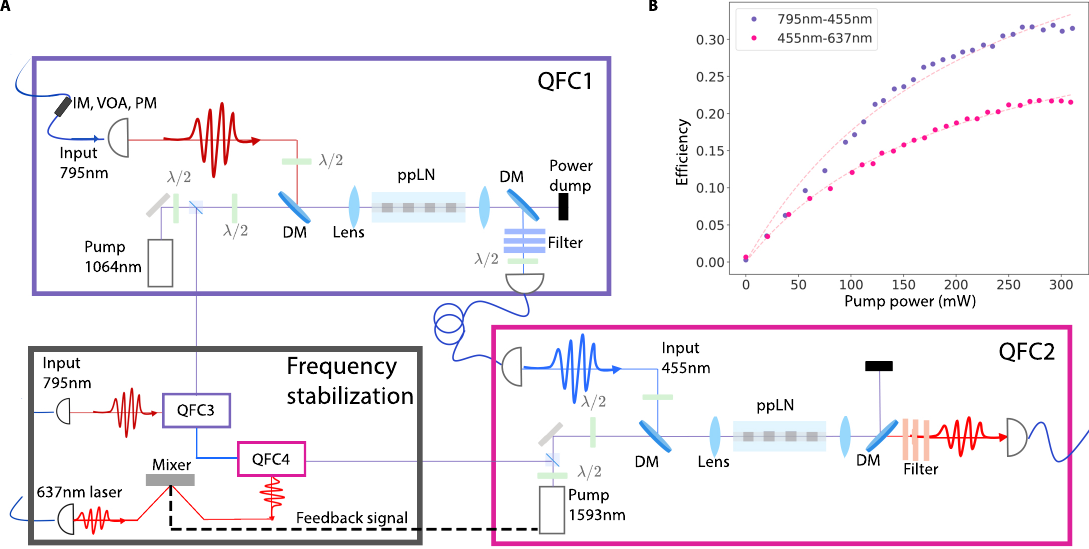}
   \caption{\textbf{Quantum frequency conversion setup.} a) Overview of the frequency conversion setup. To generate converted weak-coherent states at 637nm, the input light at 795nm undergoes two-step frequency conversion (QFC1 and QFC2), after passing through an intensity modulator (IM) to obtain the typical decay time-shape of the NV spontaneous emission, a variable optical attenuator (VOA) to manipulate the mean photon number, and a phase modulator (PM) for imprinting a phase on the time-bin qubits (only for quantum teleportation experiments). At the output of each conversion step, a dichroic mirror (DM) and a set of filters suppress residual unconverted light and pump light. A copy of the two-step frequency conversion setup (QFC3 and QFC4) is used for frequency stabilization. A higher power tap-off from the 795nm laser is converted and the resulting 637nm light is mixed with the light coming from the excitation laser of the NV. An error signal is computed and fed back to the frequency modulator of the 1593nm pump laser to match the converted light to the excitation wavelength of the NV. b) Measured efficiency for each step of the conversion while sweeping the power of the corresponding pump laser. The dashed lines represent the respective fit of the data points to a saturation curve, to extrapolate the optimal pump power. The relative error on each data point is 1\%.
   \label{fig2} 
}
\end{figure*}

To achieve wavelength indistinguishability, we employ a low-noise two-stage quantum frequency conversion process \cite{kumar_quantum_1990}, depicted in Fig. \ref{fig2}a. In the first step, the 795nm shaped weak-coherent state is overlapped with a 1064nm pump laser and coupled into a temperature-stabilized periodically-poled Lithium Niobate (ppLN) waveguide crystal, generating 455nm light via a sum-frequency conversion process with conversion efficiency of 32\% (Fig. \ref{fig2}b), measured free-space to free-space between the output of the input fiber and before the coupling into the output fiber. Subsequently, the 455nm light is down-converted to 637nm using a 1596nm pump laser, with conversion efficiency of 22\%. At the output of each ppLN crystal we include dichroic mirrors and filters to remove residual unconverted light and pump light. The overall process efficiency including in- and outcoupling and filtering is 3\%, which is sufficient for the current proof-of-concept but should be further improved in future designs. Importantly, having both pump lasers red-detuned from the signal photons results in a negligible amount of added noise in the conversion stages. To ensure that the converted light precisely matches the NV photon frequency, despite unavoidable component drifts, the frequency of converted 795nm light is locked to the NV excitation laser light. To this end, an identical two-stage conversion setup is employed with 1~mW at the input derived from the same 795nm source (Fig.~\ref{fig2}a). Details on the frequency locking procedure and the employed electronics are discussed in the Supplementary Information. The resulting spread of the beat signal is 75~kHz, pushing the corresponding contribution to teleportation infidelity well below 1\% (see below).

\section{Results}
\subsection{Two-photon quantum interference between converted 795nm photons and NV center photons}

To investigate the degree of indistinguishability of the NV photons and the converted 795nm photons, we perform a two-photon quantum interference (TPQI) experiment, also known as Hong-Ou-Mandel interference. Perfectly indistinguishable photons interfering on a balanced beam-splitter show bosonic coalescence leading to zero probability of detecting photons in both output ports of the beam-splitter~\cite{hong_measurement_1987}. 
In such an experiment, on one side we employ the NV center in its negatively charged state NV$^-$. The ground state of NV$^-$ is a spin-1 system whose spin sublevels are split by the zero-field splitting and the applied magnetic field of 25.3~mT \cite{doherty_nitrogen-vacancy_2013}. We employ the m$_S$= 0 (-1) spin state as the $\ket{0}$ ($\ket{1}$) qubit state. The NV optical transitions are spin-dependent, allowing for spin-selective optical excitation and photon emission. In the current work, we use the cycling transition $\ket{0}\rightarrow\ket{g}$, where $\ket{g}$ represent the $\ket{E_{x}}$ excited state. In the TPQI experiment, the  NV center can be modeled as a single-photon source parametrized by the probability of a photon detection per optical excitation $p_{NV}$
(counts per shot).

On the other side, the 795nm photonic states constitute a multi-photon source, featuring Poissonian photon statistics. Up to the second order, the emission probability can be approximated through the mean-photon number $|\alpha|^2$ as $|\alpha|^2+1/2|\alpha|^4$ \cite{loudon_quantum_1983}. The consequences of having two photonic sources with different statistics are discussed in detail in the SI. 

\begin{figure*}
  \includegraphics[width=1.0\linewidth]{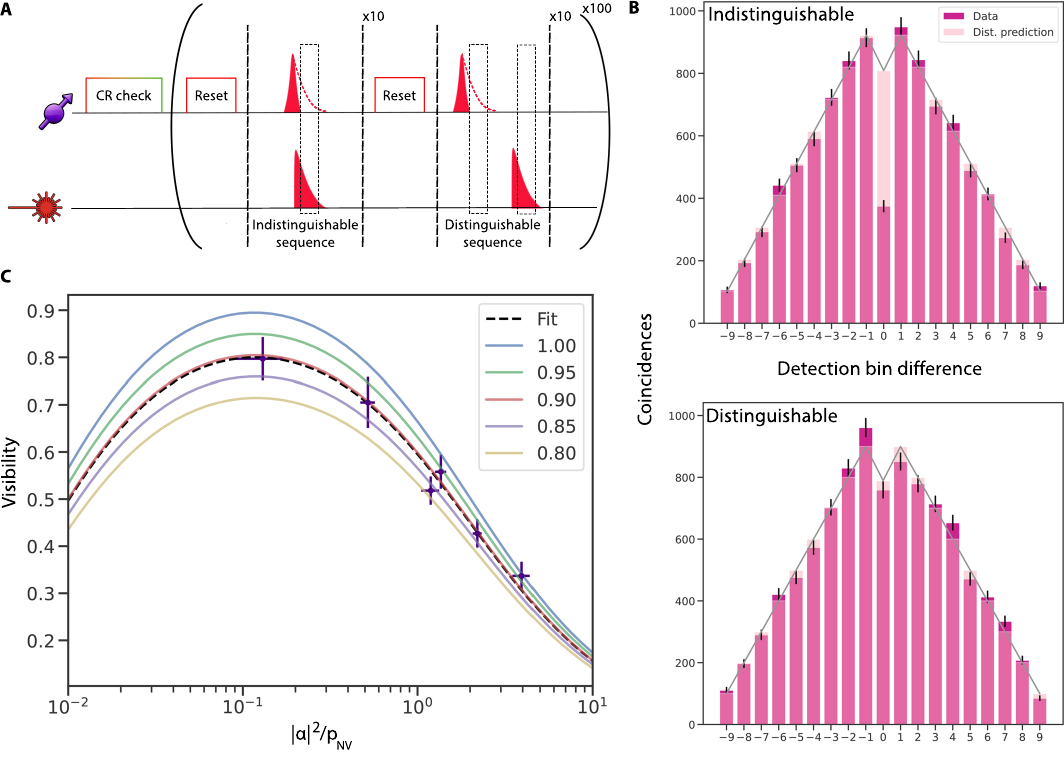}
   \caption{\textbf{Two-photon quantum interference.} a) Experimental sequence for the two-photon quantum interference experiment. The top line refers to the NV, while the bottom refers to the 795nm weak-coherent state. b) Histograms of the coincident clicks in the two sequences (magenta) for the ratio $x=|\alpha|^2$/p$_{NV}$ of 1.19$\pm$0.14. Each bar of the histograms represents coincident clicks in the two detectors within 30ns windows for all the possible combinations of a given time bin difference. As a reference, we also include, in both histograms, the expected coincidences in the case of perfectly distinguishable photons (in pink). In both diagrams, the grey line connects the expected values of the distinguishable prediction.
   In the distinguishable case, we consider two pulse windows: one around the NV photons and one around the converted 795nm photonic states. Therefore, the histogram contains the contribution of coincident clicks for three possible cases: coincident clicks in the NV window, in the converted 795nm window and in the combined windows. c) Extracted visibility for different values of $x$. The values are fitted according to the visibility model included in the SI. The dashed line represents the fit result, corresponding to indistinguishability of 0.895$\pm$0.019, while the colored lines represent our model of the visibility for different values of indistinguishability.
   \label{fig3} 
}
\end{figure*}

In Fig. \ref{fig3}a, the experimental sequence for the TPQI experiment is depicted. In the first step, a Charge-Resonance (CR) check is performed \cite{bernien_heralded_2013}, which ensures that the NV center is in the correct charge state (namely, NV$^-$) and the lasers are on resonance with the relevant NV transitions. When the CR check threshold is satisfied, the actual TPQI experiment is triggered. Two trains of 10 optical $\pi$-pulses each, which we define as 10 different bins, are sent to the NV, which leads to 20 possible emission windows  (10 per train). Each train is preceded by an optical spin-reset pulse that prepares the NV in the $\ket{0}$ state. In parallel, two trains of 10 decay-shaped pulses each are sent from the 795nm laser. The mean-photon number can be manipuleted via a variable optical attenuator (VOA). As illustrated in Fig. \ref{fig3}a, the first train of pulses constitutes the indistinguishable sequence, with the two photonic states overlapping in time on the beam splitter. The second train is the distinguishable sequence: each 795nm photonic state is delayed by 50ns with respect to the corresponding NV photon, rendering the photons fully distinguishable. The sequence of two trains is repeated 100 times before returning to the CR check. The next CR check validates both the previous TPQI sequence and, in case the threshold is satisfied, directly triggers the next sequence. The emitted photons from both sides impinge on a 50:50 in-fiber beam splitter, whose output ports are connected to two single-photon detectors. A timetagger registers the detection times of the photons in the two output ports, enabling the reconstruction of the histograms in Fig. 3b. Each bin of the histogram counts the number of coincident clicks in the two beam splitter output arms for the respective bin differences.

From the histograms in Fig. \ref{fig3}b we extract the visibility $V$=1-$\frac{p_{ind}}{p_{dist}}$, where $p_{ind}$ ($p_{dist}$) is the probability of a coincidence detection if the photons are indistinguishable (distinguishable) in the 0-bin difference. Taking into account that the photonic states follow different statistics and introducing the indistinguishability $\eta$, the visibility can be expressed as $V=\\
\dfrac{\eta x}{\dfrac{1}{2}g^{(2)}(0) + \dfrac{1}{2}x^2 + x + \dfrac{2 p_{noise} (1+x)}{p_{NV}} + \dfrac{2 p_{noise}^2}{p_{NV}^2}}$ \\
(see SI for the derivation), where $x$ is the ratio $|\alpha|^2/p_{NV}$ and $p_{noise}$ is the probability of a background (noise) click per 30ns window in one detector.

By performing this TPQI experiment and extracting the visibility for 6 different values of $x$, we can reconstruct the visibility function as shown in Fig. \ref{fig3}c. We also plot the expected visibility function, using the independently measured values for the NV $g^{(2)}$ of 0.011$\pm$0.004 and $p_{NV}$ of (5.76$\pm$0.20)e$^{-4}$, for several values of $\eta$. The value of p$_{noise}$ is discussed in the SI. We observe that the data follows the model closely over the full range. From a fit to the data we obtain the indistinguishability $\eta$=(0.895$\pm$0.019), showing that we have matched all the relevant degrees of freedom of the two photonic states to a high level.

\subsection{Qubit teleportation from a 795nm photonic time-bin qubit to the NV center spin qubit}
\begin{figure*}[htp]
  \includegraphics{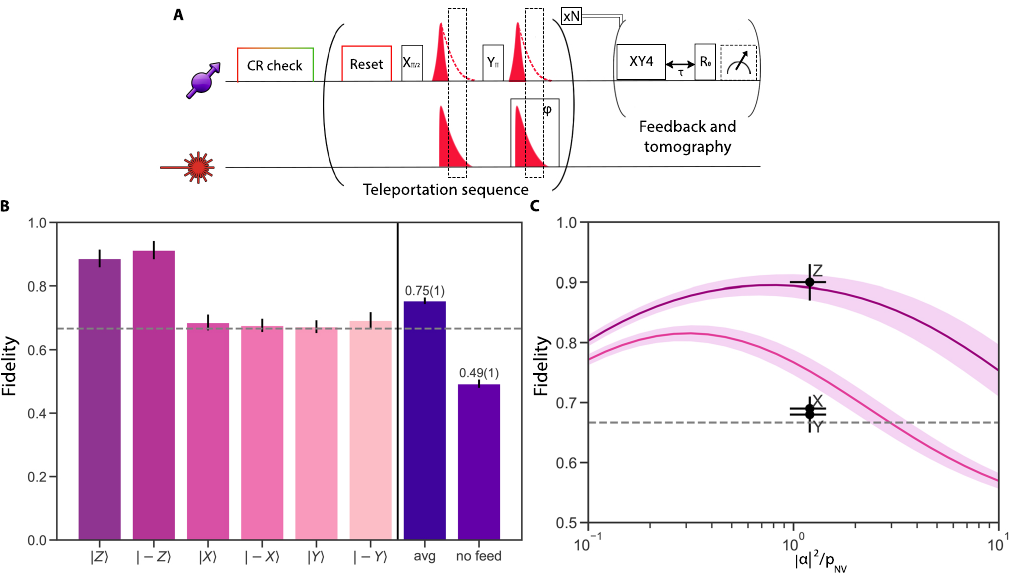}
   \caption{\textbf{Quantum teleportation of a time-bin qubit into the electron spin of the NV center.}
    a) Experimental sequence. After passing the CR check, the AWG plays the teleportation sequence as described in the main text. Such a sequence is played $N$ times. The timeout for the teleportation sequence is set to 50 repetitions, after which the NV center goes back to CR check. If a valid click pattern is detected by the FPGA in the Bell-state measurement, the AWG jumps out from the teleportation sequence and starts the feedback and tomography sequence. This sequence contains an XY4 set of pulses, where the first pulse is played after a time $\tau$ with respect to the $\pi$/2 rotation pulse in the teleportation sequence. The value $\tau$ is a multiple of the Larmor period of the electron spin, optimized taking into account the effects of the spin bath. After the XY4 sequence, the base selection pulse is played taking into account the input state and the Bell-state measurement outcome. Finally, the tomography single-shot readout of the NV electron spin is performed at the MCU level. b) Histogram showing the individual fidelities per each cardinal state, as well as the average and the resulting fidelity when no feedback operations are applied to the NV electron spin qubit. The results are corrected for tomography errors, but not for preparation errors. c) Simulation curves based on the model described in the SI. The data points in black represent the average fidelity for the states along the three axes of the Bloch sphere. The dashed line indicates the corrected classical bound.   
}
\label{fig:fig4}
\end{figure*}

Having established the high indistinguishability of the photonic states involved, we exploit the interface to perform quantum teleportation of 795nm time-bin qubits to the NV electron spin qubit, as illustrated in the diagram in Fig. \ref{fig:fig4}a. Real-time feed-forward is included to complete the teleportation, enabling the correction of phase-flipped outcomes in the Bell-state measurement and delivery of the teleported state ``alive".

The 795nm time-bin qubit is constituted by an early and late weak-coherent state separated by 300~ns and generated in the same way as in the TPQI experiment. Additionally, we include a phase modulator (PM) to manipulate the phase difference between the early and late temporal modes. The resulting qubit state is therefore in the general form of $\alpha\ket{E}+e^{i\theta}\beta\ket{L})$ with $\ket{E}$ ($\ket{L}$) denoting the early (late) time bin. For this experiment, we prepare time-bin qubits in an unbiased set of states (the cardinal states) that we indicate as: $\ket{Z}$, $\ket{-Z}$, $\ket{X}$, $\ket{-X}$, $\ket{Y}$, $\ket{-Y}$, referring to their position on the Bloch sphere.

On the NV's side, the electron spin qubit is optically initialized in $\ket{0}$. A microwave $\pi/2$ rotation along $\hat{x}$ axis of the Bloch sphere brings the qubit into a balanced superposition state. An optical $\pi$-pulse excites the NV's population in $\ket{0}$, enabling the spontaneous emission of a photon in the early time bin. Subsequently, the electron spin goes through a microwave $\pi$ rotation along $\hat{y}$ axis, and another optical $\pi$-~pulse enables the NV to spontaneously emit the late time-bin photon. The resulting NV-photon entangled state is $1/\sqrt2(|1\rangle | E\rangle \pm |0\rangle |L\rangle)$. Throughout the teleportation experiment we keep the ratio $|\alpha|^2$/p$_{NV}$ constant at 1.20$\pm$0.24 by regular recalibration.

The converted 795nm photonic state and the NV photon interfere on the balanced beam splitter, erasing the which-path information. Successful teleportation is heralded by the detection of a photon in each of the two time-bins. We can discriminate between the Bell states $\ket{\Psi^+}$ and $\ket{\Psi^-}$ by the double-click pattern: two clicks on the same detector for $\ket{\Psi^+}$ and two clicks on two different detectors for $\ket{\Psi^-}$. The valid detector clicks are detected by an FPGA in a 50ns window around the corresponding photons' time of arrival. In data analysis, we further shorten the valid teleportation time window to 20ns for an improved signal-to-noise ratio. The teleportation sequence is repeated for a maximum of 50 times before going back to the CR check.

When the FPGA detects a valid click pattern, it sends a two-bit message to the AWG. The AWG jumps out from the teleportation attempt sequence (Fig. \ref{fig:fig4}a) and starts the feedback and tomography sequence for the electron spin state. This sequence is composed of an XY4 dynamical decoupling sequence~\cite{wang_comparison_2012} followed by a basis selection pulse for the tomography. The latter is selected in real-time, taking the detector click pattern into account by applying a phase-flip correction when necessary. Finally, a single-shot readout of the NV spin qubit is performed.
Throughout the measurement, a set of automated measurement and calibration routines detect anomalies in the converted frequency and in the reset frequency, declaring those datasets as failed when the required parameters are not met (more details in the SI).
In Fig. \ref{fig:fig4}b, the results for the teleportation of the six cardinal states are reported together with the average state fidelity. The average fidelity is obtained as $F_{avg}$=1/3$\bar{Z}$+1/3$\bar{X}$+1/3$\bar{Y}$, where $\bar{Z}$, $\bar{X}$, $\bar{Y}$ represent the average fidelity along the respective axis. The resulting fidelity of (75.5 $\pm$ 1.0)\%. is well above the classical bound, which is set taking into account the use of a multi-photon source (see SI). Additionally, we also calculate the average fidelity for the equatorial states in the absence of feed-forward. In this case, the fidelity is consistent with a fully mixed state, confirming the critical role of feedback in the teleportation protocol. In the measured fidelities, we also filter based on the CR check's validation.

In Fig. \ref{fig:fig4}c we report the comparison between the measured fidelities and the predicted values of our model as a function of the ratio $|\alpha|^2$/p$_{NV}$. The model includes the effect from leakage of the intensity modulator, resulting in a preparation error for the $Z$ states of around 4\%. More details on the simulated curves are included in the SI. The small discrepancy between the $X$ and $Y$ data points and the simulation may be due to errors not captured by the model. On one side, the model does not consider imperfections in the microwave pulses that implement the NV quantum gates, which we estimate to cause an accumulated error below 1.5\%. On the time-bin qubit side, our model does not take into account phase errors due to imperfections in the fast phase modulation, which affect the preparation of the $X$ and $Y$ states but not $Z$. Correcting for the input photonic qubit preparation errors yields a best estimate for the teleportation fidelity of (78.3 $\pm$ 0.9)\%.

\section{Discussion}
We have benchmarked a photonic interface between 795nm converted time-bin qubits and an NV center-based quantum processor. The time-bin qubits are compatible with Thulium-doped crystals employed for quantum memories as well as Rubidium gas quantum memories. The interface exhibits a high photon indistinguishability, thanks to a low-noise two-step quantum frequency conversion setup, that leads to beating the classical bound for the quantum teleportation protocol, together with the capabilities of the NV center as quantum processor, which shows long coherence time and a reliable optical interface. Additionally, the implementation of control scripts made the setup to be operable at distance and for long periods. 
Our results demonstrate the realization of interfaces between heterogeneous platforms that constitute the building-blocks of the future Quantum Internet. The interface presented in this work is versatile, as the methods and results presented can be transferred to platforms with similar functionalities. Hence, further improvements can be targeted at several aspects, like application field, experimental rate and bandwidth matching. Some examples might include the use of actual quantum memories that can be synchronized with the photon emission from the NV center, along with the integration of NV centers into optical cavities \cite{ruf_resonant_2021} for higher photon rate. Another possibility is the use of different color center defects in diamonds, like the group-IV, that promise higher photon emission rates and the possibility of integration into nanophotonic structures \cite{hepp_electronic_2014, bhaskar_quantum_2017, nguyen_quantum_2019, rugar_quantum_2021}. Other promising quantum processor platforms might include defect centers in SiC \cite{lukin_integrated_2020}, Si \cite{higginbottom_optical_2022} and optical quantum dots \cite{lodahl_quantum-dot_2017, liu_violation_2023}. Additionally, higher efficiency frequency conversion setups to telecom wavelengths \cite{schafer_two-stage_2023, bersin_telecom_2023} can be employed to convert the photons emitted from both parties, leading towards the real-case scenario of quantum networks over long distances. 
\section{Methods}
\subsection{Fidelity calculation}
The fidelity for each teleported state is calculated as $F=(1+\frac{R_{\braket{i}{i}}-R_{\braket{j}{i}}}{R_{\braket{i}{i}}+R_{\braket{j}{i}}}$)/2, given that we prepared the time-bin qubit in the state $\ket{i}$ and we measure in the state $\ket{i}$ and in its orthogonal state $\ket{j}$. The quantity $R_{\braket{}{}}$ represents the tomography-related single-shot readout outcome, including the correction for known errors (see Supplementary of \cite{pompili_realization_2021}).
\section{Acknowledgements}
We thank R.F.L. Vermeulen for designing and building the electronics for the frequency stabilization, P. Botma for help with the FPGA programming, K.L. van der Enden for experimental support, J.A. Slater for fruitful discussions. 
We acknowledge funding from the Dutch Research Council (NWO) through the Spinoza prize 2019 (project number SPI 63-264) and through the projects “QuTech Part I Fundamental research” (project number 601.QT.001-1) and “QuTech Part II Applied-oriented research” (project number 601.QT.001-2), and from the EU Flagship on Quantum Technologies through the project Quantum Internet Alliance (EU Horizon 2020, grant agreement no. 820445). 
\section{Author Contributions}
M.I., M.-C.S., A.J.S., M.J.W., W.T. and R.H. devised the experiments, M.I. and M.-C.S. carried out the experiments, collected and analyzed data and discussed the results with all authors. M.I., M.-C.S., A.J.S. and M.J.W. prepared the NV setup. M.I.,M.-C.S, M.J.W, E.L., G.C.d.A. and M.O.S. designed, built and optimized the frequency conversion setups. M.I., M.-C.S., T.C., N.A. prepared the photonic time-bin qubit platform and integrated it in the setup. 
M.I. and R.H. wrote the main manuscript with input from all the authors and M.I and M.-C.S. wrote the Supplementary Information. R.H. and W.T. supervised the research.
\section{Competing interests}
The authors declare no competing interests.
\section{Data availability}
The data and the code used to produce the figures in this manuscript are available at 4TU.ResearchData (DOI: 10.4121/2ccd7708-67b9-491b-8811-e73971169370) after peer-review is complete.

\bibliographystyle{apsrev4-2} 
\bibliography{references}

\begin{thebibliography}{50}%
\makeatletter
\providecommand \@ifxundefined [1]{%
 \@ifx{#1\undefined}
}%
\providecommand \@ifnum [1]{%
 \ifnum #1\expandafter \@firstoftwo
 \else \expandafter \@secondoftwo
 \fi
}%
\providecommand \@ifx [1]{%
 \ifx #1\expandafter \@firstoftwo
 \else \expandafter \@secondoftwo
 \fi
}%
\providecommand \natexlab [1]{#1}%
\providecommand \enquote  [1]{``#1''}%
\providecommand \bibnamefont  [1]{#1}%
\providecommand \bibfnamefont [1]{#1}%
\providecommand \citenamefont [1]{#1}%
\providecommand \href@noop [0]{\@secondoftwo}%
\providecommand \href [0]{\begingroup \@sanitize@url \@href}%
\providecommand \@href[1]{\@@startlink{#1}\@@href}%
\providecommand \@@href[1]{\endgroup#1\@@endlink}%
\providecommand \@sanitize@url [0]{\catcode `\\12\catcode `\$12\catcode
  `\&12\catcode `\#12\catcode `\^12\catcode `\_12\catcode `\%12\relax}%
\providecommand \@@startlink[1]{}%
\providecommand \@@endlink[0]{}%
\providecommand \url  [0]{\begingroup\@sanitize@url \@url }%
\providecommand \@url [1]{\endgroup\@href {#1}{\urlprefix }}%
\providecommand \urlprefix  [0]{URL }%
\providecommand \Eprint [0]{\href }%
\providecommand \doibase [0]{https://doi.org/}%
\providecommand \selectlanguage [0]{\@gobble}%
\providecommand \bibinfo  [0]{\@secondoftwo}%
\providecommand \bibfield  [0]{\@secondoftwo}%
\providecommand \translation [1]{[#1]}%
\providecommand \BibitemOpen [0]{}%
\providecommand \bibitemStop [0]{}%
\providecommand \bibitemNoStop [0]{.\EOS\space}%
\providecommand \EOS [0]{\spacefactor3000\relax}%
\providecommand \BibitemShut  [1]{\csname bibitem#1\endcsname}%
\let\auto@bib@innerbib\@empty
\bibitem [{\citenamefont {Kimble}(2008)}]{kimble_quantum_2008}%
  \BibitemOpen
  \bibfield  {author} {\bibinfo {author} {\bibfnamefont {H.~J.}\ \bibnamefont
  {Kimble}},\ }\href {https://doi.org/10.1038/nature07127} {\bibfield
  {journal} {\bibinfo  {journal} {Nature}\ }\textbf {\bibinfo {volume} {453}},\
  \bibinfo {pages} {1023} (\bibinfo {year} {2008})}\BibitemShut {NoStop}%
\bibitem [{\citenamefont {Wehner}\ \emph {et~al.}(2018)\citenamefont {Wehner},
  \citenamefont {Elkouss},\ and\ \citenamefont {Hanson}}]{wehner_quantum_2018}%
  \BibitemOpen
  \bibfield  {author} {\bibinfo {author} {\bibfnamefont {S.}~\bibnamefont
  {Wehner}}, \bibinfo {author} {\bibfnamefont {D.}~\bibnamefont {Elkouss}},\
  and\ \bibinfo {author} {\bibfnamefont {R.}~\bibnamefont {Hanson}},\
  }\bibfield  {journal} {\bibinfo  {journal} {Science}\ }\textbf {\bibinfo
  {volume} {362}},\ \href {https://doi.org/10.1126/science.aam9288}
  {10.1126/science.aam9288} (\bibinfo {year} {2018})\BibitemShut {NoStop}%
\bibitem [{\citenamefont {Togan}\ \emph {et~al.}(2010)\citenamefont {Togan},
  \citenamefont {Chu}, \citenamefont {Trifonov}, \citenamefont {Jiang},
  \citenamefont {Maze}, \citenamefont {Childress}, \citenamefont {Dutt},
  \citenamefont {Sørensen}, \citenamefont {Hemmer}, \citenamefont {Zibrov},\
  and\ \citenamefont {Lukin}}]{togan_quantum_2010}%
  \BibitemOpen
  \bibfield  {author} {\bibinfo {author} {\bibfnamefont {E.}~\bibnamefont
  {Togan}}, \bibinfo {author} {\bibfnamefont {Y.}~\bibnamefont {Chu}}, \bibinfo
  {author} {\bibfnamefont {A.~S.}\ \bibnamefont {Trifonov}}, \bibinfo {author}
  {\bibfnamefont {L.}~\bibnamefont {Jiang}}, \bibinfo {author} {\bibfnamefont
  {J.}~\bibnamefont {Maze}}, \bibinfo {author} {\bibfnamefont {L.}~\bibnamefont
  {Childress}}, \bibinfo {author} {\bibfnamefont {M.~V.~G.}\ \bibnamefont
  {Dutt}}, \bibinfo {author} {\bibfnamefont {A.~S.}\ \bibnamefont {Sørensen}},
  \bibinfo {author} {\bibfnamefont {P.~R.}\ \bibnamefont {Hemmer}}, \bibinfo
  {author} {\bibfnamefont {A.~S.}\ \bibnamefont {Zibrov}},\ and\ \bibinfo
  {author} {\bibfnamefont {M.~D.}\ \bibnamefont {Lukin}},\ }\href
  {https://doi.org/10.1038/nature09256} {\bibfield  {journal} {\bibinfo
  {journal} {Nature}\ }\textbf {\bibinfo {volume} {466}},\ \bibinfo {pages}
  {730} (\bibinfo {year} {2010})}\BibitemShut {NoStop}%
\bibitem [{\citenamefont {Lodahl}(2017)}]{lodahl_quantum-dot_2017}%
  \BibitemOpen
  \bibfield  {author} {\bibinfo {author} {\bibfnamefont {P.}~\bibnamefont
  {Lodahl}},\ }\href {https://doi.org/10.1088/2058-9565/aa91bb} {\bibfield
  {journal} {\bibinfo  {journal} {Quantum Science and Technology}\ }\textbf
  {\bibinfo {volume} {3}},\ \bibinfo {pages} {013001} (\bibinfo {year}
  {2017})}\BibitemShut {NoStop}%
\bibitem [{\citenamefont {Bock}\ \emph {et~al.}(2018)\citenamefont {Bock},
  \citenamefont {Eich}, \citenamefont {Kucera}, \citenamefont {Kreis},
  \citenamefont {Lenhard}, \citenamefont {Becher},\ and\ \citenamefont
  {Eschner}}]{bock_high-fidelity_2018}%
  \BibitemOpen
  \bibfield  {author} {\bibinfo {author} {\bibfnamefont {M.}~\bibnamefont
  {Bock}}, \bibinfo {author} {\bibfnamefont {P.}~\bibnamefont {Eich}}, \bibinfo
  {author} {\bibfnamefont {S.}~\bibnamefont {Kucera}}, \bibinfo {author}
  {\bibfnamefont {M.}~\bibnamefont {Kreis}}, \bibinfo {author} {\bibfnamefont
  {A.}~\bibnamefont {Lenhard}}, \bibinfo {author} {\bibfnamefont
  {C.}~\bibnamefont {Becher}},\ and\ \bibinfo {author} {\bibfnamefont
  {J.}~\bibnamefont {Eschner}},\ }\href
  {https://doi.org/10.1038/s41467-018-04341-2} {\bibfield  {journal} {\bibinfo
  {journal} {Nature Communications}\ }\textbf {\bibinfo {volume} {9}},\
  \bibinfo {pages} {1998} (\bibinfo {year} {2018})}\BibitemShut {NoStop}%
\bibitem [{\citenamefont {Ruf}\ \emph {et~al.}(2021{\natexlab{a}})\citenamefont
  {Ruf}, \citenamefont {Wan}, \citenamefont {Choi}, \citenamefont {Englund},\
  and\ \citenamefont {Hanson}}]{ruf_quantum_2021}%
  \BibitemOpen
  \bibfield  {author} {\bibinfo {author} {\bibfnamefont {M.}~\bibnamefont
  {Ruf}}, \bibinfo {author} {\bibfnamefont {N.~H.}\ \bibnamefont {Wan}},
  \bibinfo {author} {\bibfnamefont {H.}~\bibnamefont {Choi}}, \bibinfo {author}
  {\bibfnamefont {D.}~\bibnamefont {Englund}},\ and\ \bibinfo {author}
  {\bibfnamefont {R.}~\bibnamefont {Hanson}},\ }\href
  {https://doi.org/10.1063/5.0056534} {\bibfield  {journal} {\bibinfo
  {journal} {Journal of Applied Physics}\ }\textbf {\bibinfo {volume} {130}},\
  \bibinfo {pages} {070901} (\bibinfo {year} {2021}{\natexlab{a}})},\ \bibinfo
  {note} {publisher: American Institute of Physics}\BibitemShut {NoStop}%
\bibitem [{\citenamefont {Stas}\ \emph {et~al.}(2022)\citenamefont {Stas},
  \citenamefont {Huan}, \citenamefont {Machielse}, \citenamefont {Knall},
  \citenamefont {Suleymanzade}, \citenamefont {Pingault}, \citenamefont
  {Sutula}, \citenamefont {Ding}, \citenamefont {Knaut}, \citenamefont
  {Assumpcao}, \citenamefont {Wei}, \citenamefont {Bhaskar}, \citenamefont
  {Riedinger}, \citenamefont {Sukachev}, \citenamefont {Park}, \citenamefont
  {Lončar}, \citenamefont {Levonian},\ and\ \citenamefont
  {Lukin}}]{stas_robust_2022}%
  \BibitemOpen
  \bibfield  {author} {\bibinfo {author} {\bibfnamefont {P.-J.}\ \bibnamefont
  {Stas}}, \bibinfo {author} {\bibfnamefont {Y.~Q.}\ \bibnamefont {Huan}},
  \bibinfo {author} {\bibfnamefont {B.}~\bibnamefont {Machielse}}, \bibinfo
  {author} {\bibfnamefont {E.~N.}\ \bibnamefont {Knall}}, \bibinfo {author}
  {\bibfnamefont {A.}~\bibnamefont {Suleymanzade}}, \bibinfo {author}
  {\bibfnamefont {B.}~\bibnamefont {Pingault}}, \bibinfo {author}
  {\bibfnamefont {M.}~\bibnamefont {Sutula}}, \bibinfo {author} {\bibfnamefont
  {S.~W.}\ \bibnamefont {Ding}}, \bibinfo {author} {\bibfnamefont {C.~M.}\
  \bibnamefont {Knaut}}, \bibinfo {author} {\bibfnamefont {D.~R.}\ \bibnamefont
  {Assumpcao}}, \bibinfo {author} {\bibfnamefont {Y.-C.}\ \bibnamefont {Wei}},
  \bibinfo {author} {\bibfnamefont {M.~K.}\ \bibnamefont {Bhaskar}}, \bibinfo
  {author} {\bibfnamefont {R.}~\bibnamefont {Riedinger}}, \bibinfo {author}
  {\bibfnamefont {D.~D.}\ \bibnamefont {Sukachev}}, \bibinfo {author}
  {\bibfnamefont {H.}~\bibnamefont {Park}}, \bibinfo {author} {\bibfnamefont
  {M.}~\bibnamefont {Lončar}}, \bibinfo {author} {\bibfnamefont {D.~S.}\
  \bibnamefont {Levonian}},\ and\ \bibinfo {author} {\bibfnamefont {M.~D.}\
  \bibnamefont {Lukin}},\ }\href {http://arxiv.org/abs/2207.13128} {\bibinfo
  {title} {Robust multi-qubit quantum network node with integrated error
  detection}} (\bibinfo {year} {2022}),\ \bibinfo {note} {arXiv:2207.13128
  [quant-ph]}\BibitemShut {NoStop}%
\bibitem [{\citenamefont {Lago-Rivera}\ \emph {et~al.}(2023)\citenamefont
  {Lago-Rivera}, \citenamefont {Rakonjac}, \citenamefont {Grandi},\ and\
  \citenamefont {Riedmatten}}]{lago-rivera_long_2023}%
  \BibitemOpen
  \bibfield  {author} {\bibinfo {author} {\bibfnamefont {D.}~\bibnamefont
  {Lago-Rivera}}, \bibinfo {author} {\bibfnamefont {J.~V.}\ \bibnamefont
  {Rakonjac}}, \bibinfo {author} {\bibfnamefont {S.}~\bibnamefont {Grandi}},\
  and\ \bibinfo {author} {\bibfnamefont {H.~d.}\ \bibnamefont {Riedmatten}},\
  }\href {https://doi.org/10.1038/s41467-023-37518-5} {\bibfield  {journal}
  {\bibinfo  {journal} {Nature Communications}\ }\textbf {\bibinfo {volume}
  {14}},\ \bibinfo {pages} {1889} (\bibinfo {year} {2023})}\BibitemShut
  {NoStop}%
\bibitem [{\citenamefont {Liu}\ \emph {et~al.}(2023{\natexlab{a}})\citenamefont
  {Liu}, \citenamefont {Luo}, \citenamefont {Yu}, \citenamefont {Wang},
  \citenamefont {Wang}, \citenamefont {Hu}, \citenamefont {Li}, \citenamefont
  {Zheng}, \citenamefont {Yao}, \citenamefont {Yan}, \citenamefont {Teng},
  \citenamefont {Jiang}, \citenamefont {Liu}, \citenamefont {Xie},
  \citenamefont {Zhang}, \citenamefont {Mao}, \citenamefont {Jiang},
  \citenamefont {Zhang}, \citenamefont {Bao},\ and\ \citenamefont
  {Pan}}]{liu_multinode_2023}%
  \BibitemOpen
  \bibfield  {author} {\bibinfo {author} {\bibfnamefont {J.-L.}\ \bibnamefont
  {Liu}}, \bibinfo {author} {\bibfnamefont {X.-Y.}\ \bibnamefont {Luo}},
  \bibinfo {author} {\bibfnamefont {Y.}~\bibnamefont {Yu}}, \bibinfo {author}
  {\bibfnamefont {C.-Y.}\ \bibnamefont {Wang}}, \bibinfo {author}
  {\bibfnamefont {B.}~\bibnamefont {Wang}}, \bibinfo {author} {\bibfnamefont
  {Y.}~\bibnamefont {Hu}}, \bibinfo {author} {\bibfnamefont {J.}~\bibnamefont
  {Li}}, \bibinfo {author} {\bibfnamefont {M.-Y.}\ \bibnamefont {Zheng}},
  \bibinfo {author} {\bibfnamefont {B.}~\bibnamefont {Yao}}, \bibinfo {author}
  {\bibfnamefont {Z.}~\bibnamefont {Yan}}, \bibinfo {author} {\bibfnamefont
  {D.}~\bibnamefont {Teng}}, \bibinfo {author} {\bibfnamefont {J.-W.}\
  \bibnamefont {Jiang}}, \bibinfo {author} {\bibfnamefont {X.-B.}\ \bibnamefont
  {Liu}}, \bibinfo {author} {\bibfnamefont {X.-P.}\ \bibnamefont {Xie}},
  \bibinfo {author} {\bibfnamefont {J.}~\bibnamefont {Zhang}}, \bibinfo
  {author} {\bibfnamefont {Q.-H.}\ \bibnamefont {Mao}}, \bibinfo {author}
  {\bibfnamefont {X.}~\bibnamefont {Jiang}}, \bibinfo {author} {\bibfnamefont
  {Q.}~\bibnamefont {Zhang}}, \bibinfo {author} {\bibfnamefont {X.-H.}\
  \bibnamefont {Bao}},\ and\ \bibinfo {author} {\bibfnamefont {J.-W.}\
  \bibnamefont {Pan}},\ }\href {http://arxiv.org/abs/2309.00221} {\bibinfo
  {title} {A multinode quantum network over a metropolitan area}} (\bibinfo
  {year} {2023}{\natexlab{a}})\BibitemShut {NoStop}%
\bibitem [{\citenamefont {Pompili}\ \emph {et~al.}(2021)\citenamefont
  {Pompili}, \citenamefont {Hermans}, \citenamefont {Baier}, \citenamefont
  {Beukers}, \citenamefont {Humphreys}, \citenamefont {Schouten}, \citenamefont
  {Vermeulen}, \citenamefont {Tiggelman}, \citenamefont {dos Santos~Martins},
  \citenamefont {Dirkse}, \citenamefont {Wehner},\ and\ \citenamefont
  {Hanson}}]{pompili_realization_2021}%
  \BibitemOpen
  \bibfield  {author} {\bibinfo {author} {\bibfnamefont {M.}~\bibnamefont
  {Pompili}}, \bibinfo {author} {\bibfnamefont {S.~L.~N.}\ \bibnamefont
  {Hermans}}, \bibinfo {author} {\bibfnamefont {S.}~\bibnamefont {Baier}},
  \bibinfo {author} {\bibfnamefont {H.~K.~C.}\ \bibnamefont {Beukers}},
  \bibinfo {author} {\bibfnamefont {P.~C.}\ \bibnamefont {Humphreys}}, \bibinfo
  {author} {\bibfnamefont {R.~N.}\ \bibnamefont {Schouten}}, \bibinfo {author}
  {\bibfnamefont {R.~F.~L.}\ \bibnamefont {Vermeulen}}, \bibinfo {author}
  {\bibfnamefont {M.~J.}\ \bibnamefont {Tiggelman}}, \bibinfo {author}
  {\bibfnamefont {L.}~\bibnamefont {dos Santos~Martins}}, \bibinfo {author}
  {\bibfnamefont {B.}~\bibnamefont {Dirkse}}, \bibinfo {author} {\bibfnamefont
  {S.}~\bibnamefont {Wehner}},\ and\ \bibinfo {author} {\bibfnamefont
  {R.}~\bibnamefont {Hanson}},\ }\href
  {https://doi.org/10.1126/science.abg1919} {\bibfield  {journal} {\bibinfo
  {journal} {Science}\ }\textbf {\bibinfo {volume} {372}},\ \bibinfo {pages}
  {259} (\bibinfo {year} {2021})},\ \bibinfo {note} {publisher: American
  Association for the Advancement of Science}\BibitemShut {NoStop}%
\bibitem [{\citenamefont {Hermans}\ \emph {et~al.}(2022)\citenamefont
  {Hermans}, \citenamefont {Pompili}, \citenamefont {Beukers}, \citenamefont
  {Baier}, \citenamefont {Borregaard},\ and\ \citenamefont
  {Hanson}}]{hermans_qubit_2022}%
  \BibitemOpen
  \bibfield  {author} {\bibinfo {author} {\bibfnamefont {S.~L.~N.}\
  \bibnamefont {Hermans}}, \bibinfo {author} {\bibfnamefont {M.}~\bibnamefont
  {Pompili}}, \bibinfo {author} {\bibfnamefont {H.~K.~C.}\ \bibnamefont
  {Beukers}}, \bibinfo {author} {\bibfnamefont {S.}~\bibnamefont {Baier}},
  \bibinfo {author} {\bibfnamefont {J.}~\bibnamefont {Borregaard}},\ and\
  \bibinfo {author} {\bibfnamefont {R.}~\bibnamefont {Hanson}},\ }\href
  {https://doi.org/10.1038/s41586-022-04697-y} {\bibfield  {journal} {\bibinfo
  {journal} {Nature}\ }\textbf {\bibinfo {volume} {605}},\ \bibinfo {pages}
  {663} (\bibinfo {year} {2022})}\BibitemShut {NoStop}%
\bibitem [{\citenamefont {Doherty}\ \emph {et~al.}(2013)\citenamefont
  {Doherty}, \citenamefont {Manson}, \citenamefont {Delaney}, \citenamefont
  {Jelezko}, \citenamefont {Wrachtrup},\ and\ \citenamefont
  {Hollenberg}}]{doherty_nitrogen-vacancy_2013}%
  \BibitemOpen
  \bibfield  {author} {\bibinfo {author} {\bibfnamefont {M.~W.}\ \bibnamefont
  {Doherty}}, \bibinfo {author} {\bibfnamefont {N.~B.}\ \bibnamefont {Manson}},
  \bibinfo {author} {\bibfnamefont {P.}~\bibnamefont {Delaney}}, \bibinfo
  {author} {\bibfnamefont {F.}~\bibnamefont {Jelezko}}, \bibinfo {author}
  {\bibfnamefont {J.}~\bibnamefont {Wrachtrup}},\ and\ \bibinfo {author}
  {\bibfnamefont {L.~C.}\ \bibnamefont {Hollenberg}},\ }\href
  {https://doi.org/10.1016/j.physrep.2013.02.001} {\bibfield  {journal}
  {\bibinfo  {journal} {Physics Reports}\ }\textbf {\bibinfo {volume} {528}},\
  \bibinfo {pages} {1} (\bibinfo {year} {2013})}\BibitemShut {NoStop}%
\bibitem [{\citenamefont {Childress}\ and\ \citenamefont
  {Hanson}(2013)}]{childress_diamond_2013}%
  \BibitemOpen
  \bibfield  {author} {\bibinfo {author} {\bibfnamefont {L.}~\bibnamefont
  {Childress}}\ and\ \bibinfo {author} {\bibfnamefont {R.}~\bibnamefont
  {Hanson}},\ }\href {https://doi.org/10.1557/mrs.2013.20} {\bibfield
  {journal} {\bibinfo  {journal} {MRS Bulletin}\ }\textbf {\bibinfo {volume}
  {38}},\ \bibinfo {pages} {134} (\bibinfo {year} {2013})}\BibitemShut
  {NoStop}%
\bibitem [{\citenamefont {Thiel}\ \emph {et~al.}(2014)\citenamefont {Thiel},
  \citenamefont {Sinclair}, \citenamefont {Tittel},\ and\ \citenamefont
  {Cone}}]{thiel_tm3y3ga5o12_2014}%
  \BibitemOpen
  \bibfield  {author} {\bibinfo {author} {\bibfnamefont {C.}~\bibnamefont
  {Thiel}}, \bibinfo {author} {\bibfnamefont {N.}~\bibnamefont {Sinclair}},
  \bibinfo {author} {\bibfnamefont {W.}~\bibnamefont {Tittel}},\ and\ \bibinfo
  {author} {\bibfnamefont {R.}~\bibnamefont {Cone}},\ }\href
  {https://doi.org/10.1103/PhysRevLett.113.160501} {\bibfield  {journal}
  {\bibinfo  {journal} {Physical Review Letters}\ }\textbf {\bibinfo {volume}
  {113}},\ \bibinfo {pages} {160501} (\bibinfo {year} {2014})}\BibitemShut
  {NoStop}%
\bibitem [{\citenamefont {Sinclair}\ \emph {et~al.}(2016)\citenamefont
  {Sinclair}, \citenamefont {Heshami}, \citenamefont {Deshmukh}, \citenamefont
  {Oblak}, \citenamefont {Simon},\ and\ \citenamefont
  {Tittel}}]{sinclair_proposal_2016}%
  \BibitemOpen
  \bibfield  {author} {\bibinfo {author} {\bibfnamefont {N.}~\bibnamefont
  {Sinclair}}, \bibinfo {author} {\bibfnamefont {K.}~\bibnamefont {Heshami}},
  \bibinfo {author} {\bibfnamefont {C.}~\bibnamefont {Deshmukh}}, \bibinfo
  {author} {\bibfnamefont {D.}~\bibnamefont {Oblak}}, \bibinfo {author}
  {\bibfnamefont {C.}~\bibnamefont {Simon}},\ and\ \bibinfo {author}
  {\bibfnamefont {W.}~\bibnamefont {Tittel}},\ }\href
  {https://doi.org/10.1038/ncomms13454} {\bibfield  {journal} {\bibinfo
  {journal} {Nature Communications}\ }\textbf {\bibinfo {volume} {7}},\
  \bibinfo {pages} {13454} (\bibinfo {year} {2016})}\BibitemShut {NoStop}%
\bibitem [{\citenamefont {Askarani}\ \emph {et~al.}(2021)\citenamefont
  {Askarani}, \citenamefont {Das}, \citenamefont {Davidson}, \citenamefont
  {Amaral}, \citenamefont {Sinclair}, \citenamefont {Slater}, \citenamefont
  {Marzban}, \citenamefont {Thiel}, \citenamefont {Cone}, \citenamefont
  {Oblak},\ and\ \citenamefont {Tittel}}]{askarani_long-lived_2021}%
  \BibitemOpen
  \bibfield  {author} {\bibinfo {author} {\bibfnamefont {M.~F.}\ \bibnamefont
  {Askarani}}, \bibinfo {author} {\bibfnamefont {A.}~\bibnamefont {Das}},
  \bibinfo {author} {\bibfnamefont {J.~H.}\ \bibnamefont {Davidson}}, \bibinfo
  {author} {\bibfnamefont {G.~C.}\ \bibnamefont {Amaral}}, \bibinfo {author}
  {\bibfnamefont {N.}~\bibnamefont {Sinclair}}, \bibinfo {author}
  {\bibfnamefont {J.~A.}\ \bibnamefont {Slater}}, \bibinfo {author}
  {\bibfnamefont {S.}~\bibnamefont {Marzban}}, \bibinfo {author} {\bibfnamefont
  {C.~W.}\ \bibnamefont {Thiel}}, \bibinfo {author} {\bibfnamefont {R.~L.}\
  \bibnamefont {Cone}}, \bibinfo {author} {\bibfnamefont {D.}~\bibnamefont
  {Oblak}},\ and\ \bibinfo {author} {\bibfnamefont {W.}~\bibnamefont
  {Tittel}},\ }\href {https://doi.org/10.1103/PhysRevLett.127.220502}
  {\bibfield  {journal} {\bibinfo  {journal} {Physical Review Letters}\
  }\textbf {\bibinfo {volume} {127}},\ \bibinfo {pages} {220502} (\bibinfo
  {year} {2021})}\BibitemShut {NoStop}%
\bibitem [{\citenamefont {Cho}\ \emph {et~al.}(2016)\citenamefont {Cho},
  \citenamefont {Campbell}, \citenamefont {Everett}, \citenamefont {Bernu},
  \citenamefont {Higginbottom}, \citenamefont {Cao}, \citenamefont {Geng},
  \citenamefont {Robins}, \citenamefont {Lam},\ and\ \citenamefont
  {Buchler}}]{cho_highly_2016}%
  \BibitemOpen
  \bibfield  {author} {\bibinfo {author} {\bibfnamefont {Y.-W.}\ \bibnamefont
  {Cho}}, \bibinfo {author} {\bibfnamefont {G.~T.}\ \bibnamefont {Campbell}},
  \bibinfo {author} {\bibfnamefont {J.~L.}\ \bibnamefont {Everett}}, \bibinfo
  {author} {\bibfnamefont {J.}~\bibnamefont {Bernu}}, \bibinfo {author}
  {\bibfnamefont {D.~B.}\ \bibnamefont {Higginbottom}}, \bibinfo {author}
  {\bibfnamefont {M.~T.}\ \bibnamefont {Cao}}, \bibinfo {author} {\bibfnamefont
  {J.}~\bibnamefont {Geng}}, \bibinfo {author} {\bibfnamefont {N.~P.}\
  \bibnamefont {Robins}}, \bibinfo {author} {\bibfnamefont {P.~K.}\
  \bibnamefont {Lam}},\ and\ \bibinfo {author} {\bibfnamefont {B.~C.}\
  \bibnamefont {Buchler}},\ }\href {https://doi.org/10.1364/OPTICA.3.000100}
  {\bibfield  {journal} {\bibinfo  {journal} {Optica}\ }\textbf {\bibinfo
  {volume} {3}},\ \bibinfo {pages} {100} (\bibinfo {year} {2016})}\BibitemShut
  {NoStop}%
\bibitem [{\citenamefont {Zhao}\ \emph {et~al.}(2009)\citenamefont {Zhao},
  \citenamefont {Dudin}, \citenamefont {Jenkins}, \citenamefont {Campbell},
  \citenamefont {Matsukevich}, \citenamefont {Kennedy},\ and\ \citenamefont
  {Kuzmich}}]{zhao_long-lived_2009}%
  \BibitemOpen
  \bibfield  {author} {\bibinfo {author} {\bibfnamefont {R.}~\bibnamefont
  {Zhao}}, \bibinfo {author} {\bibfnamefont {Y.~O.}\ \bibnamefont {Dudin}},
  \bibinfo {author} {\bibfnamefont {S.~D.}\ \bibnamefont {Jenkins}}, \bibinfo
  {author} {\bibfnamefont {C.~J.}\ \bibnamefont {Campbell}}, \bibinfo {author}
  {\bibfnamefont {D.~N.}\ \bibnamefont {Matsukevich}}, \bibinfo {author}
  {\bibfnamefont {T.~a.~B.}\ \bibnamefont {Kennedy}},\ and\ \bibinfo {author}
  {\bibfnamefont {A.}~\bibnamefont {Kuzmich}},\ }\href
  {https://doi.org/10.1038/nphys1152} {\bibfield  {journal} {\bibinfo
  {journal} {Nature Physics}\ }\textbf {\bibinfo {volume} {5}},\ \bibinfo
  {pages} {100} (\bibinfo {year} {2009})},\ \bibinfo {note} {number: 2
  Publisher: Nature Publishing Group}\BibitemShut {NoStop}%
\bibitem [{\citenamefont {Rosenfeld}\ \emph {et~al.}(2007)\citenamefont
  {Rosenfeld}, \citenamefont {Berner}, \citenamefont {Volz}, \citenamefont
  {Weber},\ and\ \citenamefont {Weinfurter}}]{rosenfeld_remote_2007}%
  \BibitemOpen
  \bibfield  {author} {\bibinfo {author} {\bibfnamefont {W.}~\bibnamefont
  {Rosenfeld}}, \bibinfo {author} {\bibfnamefont {S.}~\bibnamefont {Berner}},
  \bibinfo {author} {\bibfnamefont {J.}~\bibnamefont {Volz}}, \bibinfo {author}
  {\bibfnamefont {M.}~\bibnamefont {Weber}},\ and\ \bibinfo {author}
  {\bibfnamefont {H.}~\bibnamefont {Weinfurter}},\ }\href
  {https://doi.org/10.1103/PhysRevLett.98.050504} {\bibfield  {journal}
  {\bibinfo  {journal} {Physical Review Letters}\ }\textbf {\bibinfo {volume}
  {98}},\ \bibinfo {pages} {050504} (\bibinfo {year} {2007})}\BibitemShut
  {NoStop}%
\bibitem [{\citenamefont {Heller}\ \emph {et~al.}(2020)\citenamefont {Heller},
  \citenamefont {Farrera}, \citenamefont {Heinze},\ and\ \citenamefont
  {de~Riedmatten}}]{heller_cold-atom_2020}%
  \BibitemOpen
  \bibfield  {author} {\bibinfo {author} {\bibfnamefont {L.}~\bibnamefont
  {Heller}}, \bibinfo {author} {\bibfnamefont {P.}~\bibnamefont {Farrera}},
  \bibinfo {author} {\bibfnamefont {G.}~\bibnamefont {Heinze}},\ and\ \bibinfo
  {author} {\bibfnamefont {H.}~\bibnamefont {de~Riedmatten}},\ }\href
  {https://doi.org/10.1103/PhysRevLett.124.210504} {\bibfield  {journal}
  {\bibinfo  {journal} {Physical Review Letters}\ }\textbf {\bibinfo {volume}
  {124}},\ \bibinfo {pages} {210504} (\bibinfo {year} {2020})}\BibitemShut
  {NoStop}%
\bibitem [{\citenamefont {Azuma}\ \emph {et~al.}(2023)\citenamefont {Azuma},
  \citenamefont {Economou}, \citenamefont {Elkouss}, \citenamefont {Hilaire},
  \citenamefont {Jiang}, \citenamefont {Lo},\ and\ \citenamefont
  {Tzitrin}}]{azuma_quantum_2023}%
  \BibitemOpen
  \bibfield  {author} {\bibinfo {author} {\bibfnamefont {K.}~\bibnamefont
  {Azuma}}, \bibinfo {author} {\bibfnamefont {S.~E.}\ \bibnamefont {Economou}},
  \bibinfo {author} {\bibfnamefont {D.}~\bibnamefont {Elkouss}}, \bibinfo
  {author} {\bibfnamefont {P.}~\bibnamefont {Hilaire}}, \bibinfo {author}
  {\bibfnamefont {L.}~\bibnamefont {Jiang}}, \bibinfo {author} {\bibfnamefont
  {H.-K.}\ \bibnamefont {Lo}},\ and\ \bibinfo {author} {\bibfnamefont
  {I.}~\bibnamefont {Tzitrin}},\ }\href
  {https://doi.org/10.1103/RevModPhys.95.045006} {\bibfield  {journal}
  {\bibinfo  {journal} {Reviews of Modern Physics}\ }\textbf {\bibinfo {volume}
  {95}},\ \bibinfo {pages} {045006} (\bibinfo {year} {2023})}\BibitemShut
  {NoStop}%
\bibitem [{\citenamefont {Drmota}\ \emph {et~al.}(2023)\citenamefont {Drmota},
  \citenamefont {Nadlinger}, \citenamefont {Main}, \citenamefont {Nichol},
  \citenamefont {Ainley}, \citenamefont {Leichtle}, \citenamefont {Mantri},
  \citenamefont {Kashefi}, \citenamefont {Srinivas}, \citenamefont {Araneda},
  \citenamefont {Ballance},\ and\ \citenamefont
  {Lucas}}]{drmota_verifiable_2023}%
  \BibitemOpen
  \bibfield  {author} {\bibinfo {author} {\bibfnamefont {P.}~\bibnamefont
  {Drmota}}, \bibinfo {author} {\bibfnamefont {D.~P.}\ \bibnamefont
  {Nadlinger}}, \bibinfo {author} {\bibfnamefont {D.}~\bibnamefont {Main}},
  \bibinfo {author} {\bibfnamefont {B.~C.}\ \bibnamefont {Nichol}}, \bibinfo
  {author} {\bibfnamefont {E.~M.}\ \bibnamefont {Ainley}}, \bibinfo {author}
  {\bibfnamefont {D.}~\bibnamefont {Leichtle}}, \bibinfo {author}
  {\bibfnamefont {A.}~\bibnamefont {Mantri}}, \bibinfo {author} {\bibfnamefont
  {E.}~\bibnamefont {Kashefi}}, \bibinfo {author} {\bibfnamefont
  {R.}~\bibnamefont {Srinivas}}, \bibinfo {author} {\bibfnamefont
  {G.}~\bibnamefont {Araneda}}, \bibinfo {author} {\bibfnamefont {C.~J.}\
  \bibnamefont {Ballance}},\ and\ \bibinfo {author} {\bibfnamefont {D.~M.}\
  \bibnamefont {Lucas}},\ }\href {https://doi.org/10.48550/arXiv.2305.02936}
  {\bibinfo {title} {Verifiable blind quantum computing with trapped ions and
  single photons}} (\bibinfo {year} {2023})\BibitemShut {NoStop}%
\bibitem [{\citenamefont {Bennett}\ \emph {et~al.}(1993)\citenamefont
  {Bennett}, \citenamefont {Brassard}, \citenamefont {Crépeau}, \citenamefont
  {Jozsa}, \citenamefont {Peres},\ and\ \citenamefont
  {Wootters}}]{bennett_teleporting_1993}%
  \BibitemOpen
  \bibfield  {author} {\bibinfo {author} {\bibfnamefont {C.~H.}\ \bibnamefont
  {Bennett}}, \bibinfo {author} {\bibfnamefont {G.}~\bibnamefont {Brassard}},
  \bibinfo {author} {\bibfnamefont {C.}~\bibnamefont {Crépeau}}, \bibinfo
  {author} {\bibfnamefont {R.}~\bibnamefont {Jozsa}}, \bibinfo {author}
  {\bibfnamefont {A.}~\bibnamefont {Peres}},\ and\ \bibinfo {author}
  {\bibfnamefont {W.~K.}\ \bibnamefont {Wootters}},\ }\href
  {https://doi.org/10.1103/PhysRevLett.70.1895} {\bibfield  {journal} {\bibinfo
   {journal} {Physical Review Letters}\ }\textbf {\bibinfo {volume} {70}},\
  \bibinfo {pages} {1895} (\bibinfo {year} {1993})}\BibitemShut {NoStop}%
\bibitem [{\citenamefont {Bouwmeester}\ \emph {et~al.}(1997)\citenamefont
  {Bouwmeester}, \citenamefont {Pan}, \citenamefont {Mattle}, \citenamefont
  {Eibl}, \citenamefont {Weinfurter},\ and\ \citenamefont
  {Zeilinger}}]{bouwmeester_experimental_1997}%
  \BibitemOpen
  \bibfield  {author} {\bibinfo {author} {\bibfnamefont {D.}~\bibnamefont
  {Bouwmeester}}, \bibinfo {author} {\bibfnamefont {J.-W.}\ \bibnamefont
  {Pan}}, \bibinfo {author} {\bibfnamefont {K.}~\bibnamefont {Mattle}},
  \bibinfo {author} {\bibfnamefont {M.}~\bibnamefont {Eibl}}, \bibinfo {author}
  {\bibfnamefont {H.}~\bibnamefont {Weinfurter}},\ and\ \bibinfo {author}
  {\bibfnamefont {A.}~\bibnamefont {Zeilinger}},\ }\href
  {https://doi.org/https://doi.org/10.1038/37539} {\bibfield  {journal}
  {\bibinfo  {journal} {Nature}\ }\textbf {\bibinfo {volume} {390}},\ \bibinfo
  {pages} {575} (\bibinfo {year} {1997})}\BibitemShut {NoStop}%
\bibitem [{\citenamefont {Heshami}\ \emph {et~al.}(2016)\citenamefont
  {Heshami}, \citenamefont {England}, \citenamefont {Humphreys}, \citenamefont
  {Bustard}, \citenamefont {Acosta}, \citenamefont {Nunn},\ and\ \citenamefont
  {Sussman}}]{heshami_quantum_2016}%
  \BibitemOpen
  \bibfield  {author} {\bibinfo {author} {\bibfnamefont {K.}~\bibnamefont
  {Heshami}}, \bibinfo {author} {\bibfnamefont {D.~G.}\ \bibnamefont
  {England}}, \bibinfo {author} {\bibfnamefont {P.~C.}\ \bibnamefont
  {Humphreys}}, \bibinfo {author} {\bibfnamefont {P.~J.}\ \bibnamefont
  {Bustard}}, \bibinfo {author} {\bibfnamefont {V.~M.}\ \bibnamefont {Acosta}},
  \bibinfo {author} {\bibfnamefont {J.}~\bibnamefont {Nunn}},\ and\ \bibinfo
  {author} {\bibfnamefont {B.~J.}\ \bibnamefont {Sussman}},\ }\href
  {https://doi.org/10.1080/09500340.2016.1148212} {\bibfield  {journal}
  {\bibinfo  {journal} {Journal of Modern Optics}\ }\textbf {\bibinfo {volume}
  {63}},\ \bibinfo {pages} {2005} (\bibinfo {year} {2016})}\BibitemShut
  {NoStop}%
\bibitem [{\citenamefont {Lei}\ \emph {et~al.}(2023)\citenamefont {Lei},
  \citenamefont {Asadi}, \citenamefont {Zhong}, \citenamefont {Kuzmich},
  \citenamefont {Simon},\ and\ \citenamefont {Hosseini}}]{lei_quantum_2023}%
  \BibitemOpen
  \bibfield  {author} {\bibinfo {author} {\bibfnamefont {Y.}~\bibnamefont
  {Lei}}, \bibinfo {author} {\bibfnamefont {F.~K.}\ \bibnamefont {Asadi}},
  \bibinfo {author} {\bibfnamefont {T.}~\bibnamefont {Zhong}}, \bibinfo
  {author} {\bibfnamefont {A.}~\bibnamefont {Kuzmich}}, \bibinfo {author}
  {\bibfnamefont {C.}~\bibnamefont {Simon}},\ and\ \bibinfo {author}
  {\bibfnamefont {M.}~\bibnamefont {Hosseini}},\ }\href
  {https://doi.org/10.1364/OPTICA.493732} {\bibfield  {journal} {\bibinfo
  {journal} {Optica}\ }\textbf {\bibinfo {volume} {10}},\ \bibinfo {pages}
  {1511} (\bibinfo {year} {2023})}\BibitemShut {NoStop}%
\bibitem [{\citenamefont {Afzelius}\ \emph {et~al.}(2009)\citenamefont
  {Afzelius}, \citenamefont {Simon}, \citenamefont {de~Riedmatten},\ and\
  \citenamefont {Gisin}}]{afzelius_multimode_2009}%
  \BibitemOpen
  \bibfield  {author} {\bibinfo {author} {\bibfnamefont {M.}~\bibnamefont
  {Afzelius}}, \bibinfo {author} {\bibfnamefont {C.}~\bibnamefont {Simon}},
  \bibinfo {author} {\bibfnamefont {H.}~\bibnamefont {de~Riedmatten}},\ and\
  \bibinfo {author} {\bibfnamefont {N.}~\bibnamefont {Gisin}},\ }\href
  {https://doi.org/10.1103/PhysRevA.79.052329} {\bibfield  {journal} {\bibinfo
  {journal} {Physical Review A}\ }\textbf {\bibinfo {volume} {79}},\ \bibinfo
  {pages} {052329} (\bibinfo {year} {2009})}\BibitemShut {NoStop}%
\bibitem [{\citenamefont {Covey}\ \emph {et~al.}(2023)\citenamefont {Covey},
  \citenamefont {Weinfurter},\ and\ \citenamefont
  {Bernien}}]{covey_quantum_2023}%
  \BibitemOpen
  \bibfield  {author} {\bibinfo {author} {\bibfnamefont {J.~P.}\ \bibnamefont
  {Covey}}, \bibinfo {author} {\bibfnamefont {H.}~\bibnamefont {Weinfurter}},\
  and\ \bibinfo {author} {\bibfnamefont {H.}~\bibnamefont {Bernien}},\ }\href
  {https://doi.org/10.1038/s41534-023-00759-9} {\bibfield  {journal} {\bibinfo
  {journal} {npj Quantum Information}\ }\textbf {\bibinfo {volume} {9}},\
  \bibinfo {pages} {1} (\bibinfo {year} {2023})}\BibitemShut {NoStop}%
\bibitem [{\citenamefont {Lipka}\ \emph {et~al.}(2021)\citenamefont {Lipka},
  \citenamefont {Mazelanik}, \citenamefont {Leszczyński}, \citenamefont
  {Wasilewski},\ and\ \citenamefont
  {Parniak}}]{lipka_massively-multiplexed_2021}%
  \BibitemOpen
  \bibfield  {author} {\bibinfo {author} {\bibfnamefont {M.}~\bibnamefont
  {Lipka}}, \bibinfo {author} {\bibfnamefont {M.}~\bibnamefont {Mazelanik}},
  \bibinfo {author} {\bibfnamefont {A.}~\bibnamefont {Leszczyński}}, \bibinfo
  {author} {\bibfnamefont {W.}~\bibnamefont {Wasilewski}},\ and\ \bibinfo
  {author} {\bibfnamefont {M.}~\bibnamefont {Parniak}},\ }\href
  {https://doi.org/10.1038/s42005-021-00551-1} {\bibfield  {journal} {\bibinfo
  {journal} {Communications Physics}\ }\textbf {\bibinfo {volume} {4}},\
  \bibinfo {pages} {1} (\bibinfo {year} {2021})}\BibitemShut {NoStop}%
\bibitem [{\citenamefont {Bussieres}\ \emph {et~al.}(2013)\citenamefont
  {Bussieres}, \citenamefont {Sangouard}, \citenamefont {Afzelius},
  \citenamefont {de~Riedmatten}, \citenamefont {Simon},\ and\ \citenamefont
  {Tittel}}]{bussieres_prospective_2013}%
  \BibitemOpen
  \bibfield  {author} {\bibinfo {author} {\bibfnamefont {F.}~\bibnamefont
  {Bussieres}}, \bibinfo {author} {\bibfnamefont {N.}~\bibnamefont
  {Sangouard}}, \bibinfo {author} {\bibfnamefont {M.}~\bibnamefont {Afzelius}},
  \bibinfo {author} {\bibfnamefont {H.}~\bibnamefont {de~Riedmatten}}, \bibinfo
  {author} {\bibfnamefont {C.}~\bibnamefont {Simon}},\ and\ \bibinfo {author}
  {\bibfnamefont {W.}~\bibnamefont {Tittel}},\ }\href
  {https://doi.org/10.1080/09500340.2013.856482} {\bibfield  {journal}
  {\bibinfo  {journal} {Journal of Modern Optics}\ }\textbf {\bibinfo {volume}
  {60}},\ \bibinfo {pages} {1519} (\bibinfo {year} {2013})}\BibitemShut
  {NoStop}%
\bibitem [{\citenamefont {Sangouard}\ \emph {et~al.}(2011)\citenamefont
  {Sangouard}, \citenamefont {Simon}, \citenamefont {de~Riedmatten},\ and\
  \citenamefont {Gisin}}]{sangouard_quantum_2011}%
  \BibitemOpen
  \bibfield  {author} {\bibinfo {author} {\bibfnamefont {N.}~\bibnamefont
  {Sangouard}}, \bibinfo {author} {\bibfnamefont {C.}~\bibnamefont {Simon}},
  \bibinfo {author} {\bibfnamefont {H.}~\bibnamefont {de~Riedmatten}},\ and\
  \bibinfo {author} {\bibfnamefont {N.}~\bibnamefont {Gisin}},\ }\href
  {https://doi.org/10.1103/RevModPhys.83.33} {\bibfield  {journal} {\bibinfo
  {journal} {Reviews of Modern Physics}\ }\textbf {\bibinfo {volume} {83}},\
  \bibinfo {pages} {33} (\bibinfo {year} {2011})}\BibitemShut {NoStop}%
\bibitem [{\citenamefont {Kumar}(1990)}]{kumar_quantum_1990}%
  \BibitemOpen
  \bibfield  {author} {\bibinfo {author} {\bibfnamefont {P.}~\bibnamefont
  {Kumar}},\ }\href {https://doi.org/10.1364/OL.15.001476} {\bibfield
  {journal} {\bibinfo  {journal} {Optics Letters}\ }\textbf {\bibinfo {volume}
  {15}},\ \bibinfo {pages} {1476} (\bibinfo {year} {1990})}\BibitemShut
  {NoStop}%
\bibitem [{\citenamefont {Hong}\ \emph {et~al.}(1987)\citenamefont {Hong},
  \citenamefont {Ou},\ and\ \citenamefont {Mandel}}]{hong_measurement_1987}%
  \BibitemOpen
  \bibfield  {author} {\bibinfo {author} {\bibfnamefont {C.~K.}\ \bibnamefont
  {Hong}}, \bibinfo {author} {\bibfnamefont {Z.~Y.}\ \bibnamefont {Ou}},\ and\
  \bibinfo {author} {\bibfnamefont {L.}~\bibnamefont {Mandel}},\ }\href
  {https://doi.org/10.1103/PhysRevLett.59.2044} {\bibfield  {journal} {\bibinfo
   {journal} {Physical Review Letters}\ }\textbf {\bibinfo {volume} {59}},\
  \bibinfo {pages} {2044} (\bibinfo {year} {1987})}\BibitemShut {NoStop}%
\bibitem [{\citenamefont {Loudon}(1983)}]{loudon_quantum_1983}%
  \BibitemOpen
  \bibfield  {author} {\bibinfo {author} {\bibfnamefont {R.}~\bibnamefont
  {Loudon}},\ }\href@noop {} {\emph {\bibinfo {title} {The {Quantum} {Theory}
  of {Light}}}}\ (\bibinfo  {publisher} {Clarendon Press},\ \bibinfo {year}
  {1983})\BibitemShut {NoStop}%
\bibitem [{\citenamefont {Bernien}\ \emph {et~al.}(2013)\citenamefont
  {Bernien}, \citenamefont {Hensen}, \citenamefont {Pfaff}, \citenamefont
  {Koolstra}, \citenamefont {Blok}, \citenamefont {Robledo}, \citenamefont
  {Taminiau}, \citenamefont {Markham}, \citenamefont {Twitchen}, \citenamefont
  {Childress},\ and\ \citenamefont {Hanson}}]{bernien_heralded_2013}%
  \BibitemOpen
  \bibfield  {author} {\bibinfo {author} {\bibfnamefont {H.}~\bibnamefont
  {Bernien}}, \bibinfo {author} {\bibfnamefont {B.}~\bibnamefont {Hensen}},
  \bibinfo {author} {\bibfnamefont {W.}~\bibnamefont {Pfaff}}, \bibinfo
  {author} {\bibfnamefont {G.}~\bibnamefont {Koolstra}}, \bibinfo {author}
  {\bibfnamefont {M.~S.}\ \bibnamefont {Blok}}, \bibinfo {author}
  {\bibfnamefont {L.}~\bibnamefont {Robledo}}, \bibinfo {author} {\bibfnamefont
  {T.~H.}\ \bibnamefont {Taminiau}}, \bibinfo {author} {\bibfnamefont
  {M.}~\bibnamefont {Markham}}, \bibinfo {author} {\bibfnamefont {D.~J.}\
  \bibnamefont {Twitchen}}, \bibinfo {author} {\bibfnamefont {L.}~\bibnamefont
  {Childress}},\ and\ \bibinfo {author} {\bibfnamefont {R.}~\bibnamefont
  {Hanson}},\ }\href {https://doi.org/10.1038/nature12016} {\bibfield
  {journal} {\bibinfo  {journal} {Nature}\ }\textbf {\bibinfo {volume} {497}},\
  \bibinfo {pages} {86} (\bibinfo {year} {2013})}\BibitemShut {NoStop}%
\bibitem [{\citenamefont {Wang}\ \emph {et~al.}(2012)\citenamefont {Wang},
  \citenamefont {de~Lange}, \citenamefont {Ristè}, \citenamefont {Hanson},\
  and\ \citenamefont {Dobrovitski}}]{wang_comparison_2012}%
  \BibitemOpen
  \bibfield  {author} {\bibinfo {author} {\bibfnamefont {Z.-H.}\ \bibnamefont
  {Wang}}, \bibinfo {author} {\bibfnamefont {G.}~\bibnamefont {de~Lange}},
  \bibinfo {author} {\bibfnamefont {D.}~\bibnamefont {Ristè}}, \bibinfo
  {author} {\bibfnamefont {R.}~\bibnamefont {Hanson}},\ and\ \bibinfo {author}
  {\bibfnamefont {V.~V.}\ \bibnamefont {Dobrovitski}},\ }\href
  {https://doi.org/10.1103/PhysRevB.85.155204} {\bibfield  {journal} {\bibinfo
  {journal} {Physical Review B}\ }\textbf {\bibinfo {volume} {85}},\ \bibinfo
  {pages} {155204} (\bibinfo {year} {2012})}\BibitemShut {NoStop}%
\bibitem [{\citenamefont {Ruf}\ \emph {et~al.}(2021{\natexlab{b}})\citenamefont
  {Ruf}, \citenamefont {Weaver}, \citenamefont {van Dam},\ and\ \citenamefont
  {Hanson}}]{ruf_resonant_2021}%
  \BibitemOpen
  \bibfield  {author} {\bibinfo {author} {\bibfnamefont {M.}~\bibnamefont
  {Ruf}}, \bibinfo {author} {\bibfnamefont {M.}~\bibnamefont {Weaver}},
  \bibinfo {author} {\bibfnamefont {S.}~\bibnamefont {van Dam}},\ and\ \bibinfo
  {author} {\bibfnamefont {R.}~\bibnamefont {Hanson}},\ }\href
  {https://doi.org/10.1103/PhysRevApplied.15.024049} {\bibfield  {journal}
  {\bibinfo  {journal} {Physical Review Applied}\ }\textbf {\bibinfo {volume}
  {15}},\ \bibinfo {pages} {024049} (\bibinfo {year}
  {2021}{\natexlab{b}})}\BibitemShut {NoStop}%
\bibitem [{\citenamefont {Hepp}\ \emph {et~al.}(2014)\citenamefont {Hepp},
  \citenamefont {Müller}, \citenamefont {Waselowski}, \citenamefont {Becker},
  \citenamefont {Pingault}, \citenamefont {Sternschulte}, \citenamefont
  {Steinmüller-Nethl}, \citenamefont {Gali}, \citenamefont {Maze},
  \citenamefont {Atatüre},\ and\ \citenamefont
  {Becher}}]{hepp_electronic_2014}%
  \BibitemOpen
  \bibfield  {author} {\bibinfo {author} {\bibfnamefont {C.}~\bibnamefont
  {Hepp}}, \bibinfo {author} {\bibfnamefont {T.}~\bibnamefont {Müller}},
  \bibinfo {author} {\bibfnamefont {V.}~\bibnamefont {Waselowski}}, \bibinfo
  {author} {\bibfnamefont {J.~N.}\ \bibnamefont {Becker}}, \bibinfo {author}
  {\bibfnamefont {B.}~\bibnamefont {Pingault}}, \bibinfo {author}
  {\bibfnamefont {H.}~\bibnamefont {Sternschulte}}, \bibinfo {author}
  {\bibfnamefont {D.}~\bibnamefont {Steinmüller-Nethl}}, \bibinfo {author}
  {\bibfnamefont {A.}~\bibnamefont {Gali}}, \bibinfo {author} {\bibfnamefont
  {J.~R.}\ \bibnamefont {Maze}}, \bibinfo {author} {\bibfnamefont
  {M.}~\bibnamefont {Atatüre}},\ and\ \bibinfo {author} {\bibfnamefont
  {C.}~\bibnamefont {Becher}},\ }\href
  {https://doi.org/10.1103/PhysRevLett.112.036405} {\bibfield  {journal}
  {\bibinfo  {journal} {Physical Review Letters}\ }\textbf {\bibinfo {volume}
  {112}},\ \bibinfo {pages} {036405} (\bibinfo {year} {2014})}\BibitemShut
  {NoStop}%
\bibitem [{\citenamefont {Bhaskar}\ \emph {et~al.}(2017)\citenamefont
  {Bhaskar}, \citenamefont {Sukachev}, \citenamefont {Sipahigil}, \citenamefont
  {Evans}, \citenamefont {Burek}, \citenamefont {Nguyen}, \citenamefont
  {Rogers}, \citenamefont {Siyushev}, \citenamefont {Metsch}, \citenamefont
  {Park}, \citenamefont {Jelezko}, \citenamefont {Lončar},\ and\ \citenamefont
  {Lukin}}]{bhaskar_quantum_2017}%
  \BibitemOpen
  \bibfield  {author} {\bibinfo {author} {\bibfnamefont {M.}~\bibnamefont
  {Bhaskar}}, \bibinfo {author} {\bibfnamefont {D.}~\bibnamefont {Sukachev}},
  \bibinfo {author} {\bibfnamefont {A.}~\bibnamefont {Sipahigil}}, \bibinfo
  {author} {\bibfnamefont {R.}~\bibnamefont {Evans}}, \bibinfo {author}
  {\bibfnamefont {M.}~\bibnamefont {Burek}}, \bibinfo {author} {\bibfnamefont
  {C.}~\bibnamefont {Nguyen}}, \bibinfo {author} {\bibfnamefont
  {L.}~\bibnamefont {Rogers}}, \bibinfo {author} {\bibfnamefont
  {P.}~\bibnamefont {Siyushev}}, \bibinfo {author} {\bibfnamefont
  {M.}~\bibnamefont {Metsch}}, \bibinfo {author} {\bibfnamefont
  {H.}~\bibnamefont {Park}}, \bibinfo {author} {\bibfnamefont {F.}~\bibnamefont
  {Jelezko}}, \bibinfo {author} {\bibfnamefont {M.}~\bibnamefont {Lončar}},\
  and\ \bibinfo {author} {\bibfnamefont {M.}~\bibnamefont {Lukin}},\ }\href
  {https://doi.org/10.1103/PhysRevLett.118.223603} {\bibfield  {journal}
  {\bibinfo  {journal} {Physical Review Letters}\ }\textbf {\bibinfo {volume}
  {118}},\ \bibinfo {pages} {223603} (\bibinfo {year} {2017})}\BibitemShut
  {NoStop}%
\bibitem [{\citenamefont {Nguyen}\ \emph {et~al.}(2019)\citenamefont {Nguyen},
  \citenamefont {Sukachev}, \citenamefont {Bhaskar}, \citenamefont {Machielse},
  \citenamefont {Levonian}, \citenamefont {Knall}, \citenamefont {Stroganov},
  \citenamefont {Riedinger}, \citenamefont {Park}, \citenamefont {Lončar},\
  and\ \citenamefont {Lukin}}]{nguyen_quantum_2019}%
  \BibitemOpen
  \bibfield  {author} {\bibinfo {author} {\bibfnamefont {C.}~\bibnamefont
  {Nguyen}}, \bibinfo {author} {\bibfnamefont {D.}~\bibnamefont {Sukachev}},
  \bibinfo {author} {\bibfnamefont {M.}~\bibnamefont {Bhaskar}}, \bibinfo
  {author} {\bibfnamefont {B.}~\bibnamefont {Machielse}}, \bibinfo {author}
  {\bibfnamefont {D.}~\bibnamefont {Levonian}}, \bibinfo {author}
  {\bibfnamefont {E.}~\bibnamefont {Knall}}, \bibinfo {author} {\bibfnamefont
  {P.}~\bibnamefont {Stroganov}}, \bibinfo {author} {\bibfnamefont
  {R.}~\bibnamefont {Riedinger}}, \bibinfo {author} {\bibfnamefont
  {H.}~\bibnamefont {Park}}, \bibinfo {author} {\bibfnamefont {M.}~\bibnamefont
  {Lončar}},\ and\ \bibinfo {author} {\bibfnamefont {M.}~\bibnamefont
  {Lukin}},\ }\href {https://doi.org/10.1103/PhysRevLett.123.183602} {\bibfield
   {journal} {\bibinfo  {journal} {Physical Review Letters}\ }\textbf {\bibinfo
  {volume} {123}},\ \bibinfo {pages} {183602} (\bibinfo {year}
  {2019})}\BibitemShut {NoStop}%
\bibitem [{\citenamefont {Rugar}\ \emph {et~al.}(2021)\citenamefont {Rugar},
  \citenamefont {Aghaeimeibodi}, \citenamefont {Riedel}, \citenamefont {Dory},
  \citenamefont {Lu}, \citenamefont {McQuade}, \citenamefont {Shen},
  \citenamefont {Melosh},\ and\ \citenamefont
  {Vučković}}]{rugar_quantum_2021}%
  \BibitemOpen
  \bibfield  {author} {\bibinfo {author} {\bibfnamefont {A.~E.}\ \bibnamefont
  {Rugar}}, \bibinfo {author} {\bibfnamefont {S.}~\bibnamefont
  {Aghaeimeibodi}}, \bibinfo {author} {\bibfnamefont {D.}~\bibnamefont
  {Riedel}}, \bibinfo {author} {\bibfnamefont {C.}~\bibnamefont {Dory}},
  \bibinfo {author} {\bibfnamefont {H.}~\bibnamefont {Lu}}, \bibinfo {author}
  {\bibfnamefont {P.~J.}\ \bibnamefont {McQuade}}, \bibinfo {author}
  {\bibfnamefont {Z.-X.}\ \bibnamefont {Shen}}, \bibinfo {author}
  {\bibfnamefont {N.~A.}\ \bibnamefont {Melosh}},\ and\ \bibinfo {author}
  {\bibfnamefont {J.}~\bibnamefont {Vučković}},\ }\href
  {https://doi.org/10.1103/PhysRevX.11.031021} {\bibfield  {journal} {\bibinfo
  {journal} {Physical Review X}\ }\textbf {\bibinfo {volume} {11}},\ \bibinfo
  {pages} {031021} (\bibinfo {year} {2021})}\BibitemShut {NoStop}%
\bibitem [{\citenamefont {Lukin}\ \emph {et~al.}(2020)\citenamefont {Lukin},
  \citenamefont {Guidry},\ and\ \citenamefont
  {Vučković}}]{lukin_integrated_2020}%
  \BibitemOpen
  \bibfield  {author} {\bibinfo {author} {\bibfnamefont {D.~M.}\ \bibnamefont
  {Lukin}}, \bibinfo {author} {\bibfnamefont {M.~A.}\ \bibnamefont {Guidry}},\
  and\ \bibinfo {author} {\bibfnamefont {J.}~\bibnamefont {Vučković}},\
  }\href {https://doi.org/10.1103/PRXQuantum.1.020102} {\bibfield  {journal}
  {\bibinfo  {journal} {PRX Quantum}\ }\textbf {\bibinfo {volume} {1}},\
  \bibinfo {pages} {020102} (\bibinfo {year} {2020})}\BibitemShut {NoStop}%
\bibitem [{\citenamefont {Higginbottom}\ \emph {et~al.}(2022)\citenamefont
  {Higginbottom}, \citenamefont {Kurkjian}, \citenamefont {Chartrand},
  \citenamefont {Kazemi}, \citenamefont {Brunelle}, \citenamefont {MacQuarrie},
  \citenamefont {Klein}, \citenamefont {Lee-Hone}, \citenamefont {Stacho},
  \citenamefont {Ruether}, \citenamefont {Bowness}, \citenamefont {Bergeron},
  \citenamefont {DeAbreu}, \citenamefont {Harrigan}, \citenamefont
  {Kanaganayagam}, \citenamefont {Marsden}, \citenamefont {Richards},
  \citenamefont {Stott}, \citenamefont {Roorda}, \citenamefont {Morse},
  \citenamefont {Thewalt},\ and\ \citenamefont
  {Simmons}}]{higginbottom_optical_2022}%
  \BibitemOpen
  \bibfield  {author} {\bibinfo {author} {\bibfnamefont {D.~B.}\ \bibnamefont
  {Higginbottom}}, \bibinfo {author} {\bibfnamefont {A.~T.~K.}\ \bibnamefont
  {Kurkjian}}, \bibinfo {author} {\bibfnamefont {C.}~\bibnamefont {Chartrand}},
  \bibinfo {author} {\bibfnamefont {M.}~\bibnamefont {Kazemi}}, \bibinfo
  {author} {\bibfnamefont {N.~A.}\ \bibnamefont {Brunelle}}, \bibinfo {author}
  {\bibfnamefont {E.~R.}\ \bibnamefont {MacQuarrie}}, \bibinfo {author}
  {\bibfnamefont {J.~R.}\ \bibnamefont {Klein}}, \bibinfo {author}
  {\bibfnamefont {N.~R.}\ \bibnamefont {Lee-Hone}}, \bibinfo {author}
  {\bibfnamefont {J.}~\bibnamefont {Stacho}}, \bibinfo {author} {\bibfnamefont
  {M.}~\bibnamefont {Ruether}}, \bibinfo {author} {\bibfnamefont
  {C.}~\bibnamefont {Bowness}}, \bibinfo {author} {\bibfnamefont
  {L.}~\bibnamefont {Bergeron}}, \bibinfo {author} {\bibfnamefont
  {A.}~\bibnamefont {DeAbreu}}, \bibinfo {author} {\bibfnamefont {S.~R.}\
  \bibnamefont {Harrigan}}, \bibinfo {author} {\bibfnamefont {J.}~\bibnamefont
  {Kanaganayagam}}, \bibinfo {author} {\bibfnamefont {D.~W.}\ \bibnamefont
  {Marsden}}, \bibinfo {author} {\bibfnamefont {T.~S.}\ \bibnamefont
  {Richards}}, \bibinfo {author} {\bibfnamefont {L.~A.}\ \bibnamefont {Stott}},
  \bibinfo {author} {\bibfnamefont {S.}~\bibnamefont {Roorda}}, \bibinfo
  {author} {\bibfnamefont {K.~J.}\ \bibnamefont {Morse}}, \bibinfo {author}
  {\bibfnamefont {M.~L.~W.}\ \bibnamefont {Thewalt}},\ and\ \bibinfo {author}
  {\bibfnamefont {S.}~\bibnamefont {Simmons}},\ }\href
  {https://doi.org/10.1038/s41586-022-04821-y} {\bibfield  {journal} {\bibinfo
  {journal} {Nature}\ }\textbf {\bibinfo {volume} {607}},\ \bibinfo {pages}
  {266} (\bibinfo {year} {2022})}\BibitemShut {NoStop}%
\bibitem [{\citenamefont {Liu}\ \emph {et~al.}(2023{\natexlab{b}})\citenamefont
  {Liu}, \citenamefont {Sandberg}, \citenamefont {Chan}, \citenamefont
  {Schrinski}, \citenamefont {Anyfantaki}, \citenamefont {Nielsen},
  \citenamefont {Larsen}, \citenamefont {Skalkin}, \citenamefont {Wang},
  \citenamefont {Midolo}, \citenamefont {Scholz}, \citenamefont {Wieck},
  \citenamefont {Ludwig}, \citenamefont {Sørensen}, \citenamefont {Tiranov},\
  and\ \citenamefont {Lodahl}}]{liu_violation_2023}%
  \BibitemOpen
  \bibfield  {author} {\bibinfo {author} {\bibfnamefont {S.}~\bibnamefont
  {Liu}}, \bibinfo {author} {\bibfnamefont {O.~A.~D.}\ \bibnamefont
  {Sandberg}}, \bibinfo {author} {\bibfnamefont {M.~L.}\ \bibnamefont {Chan}},
  \bibinfo {author} {\bibfnamefont {B.}~\bibnamefont {Schrinski}}, \bibinfo
  {author} {\bibfnamefont {Y.}~\bibnamefont {Anyfantaki}}, \bibinfo {author}
  {\bibfnamefont {R.~B.}\ \bibnamefont {Nielsen}}, \bibinfo {author}
  {\bibfnamefont {R.~G.}\ \bibnamefont {Larsen}}, \bibinfo {author}
  {\bibfnamefont {A.}~\bibnamefont {Skalkin}}, \bibinfo {author} {\bibfnamefont
  {Y.}~\bibnamefont {Wang}}, \bibinfo {author} {\bibfnamefont {L.}~\bibnamefont
  {Midolo}}, \bibinfo {author} {\bibfnamefont {S.}~\bibnamefont {Scholz}},
  \bibinfo {author} {\bibfnamefont {A.~D.}\ \bibnamefont {Wieck}}, \bibinfo
  {author} {\bibfnamefont {A.}~\bibnamefont {Ludwig}}, \bibinfo {author}
  {\bibfnamefont {A.~S.}\ \bibnamefont {Sørensen}}, \bibinfo {author}
  {\bibfnamefont {A.}~\bibnamefont {Tiranov}},\ and\ \bibinfo {author}
  {\bibfnamefont {P.}~\bibnamefont {Lodahl}},\ }\href
  {https://doi.org/10.48550/arXiv.2306.12801} {\bibinfo {title} {Violation of
  {Bell} inequality by photon scattering on a two-level emitter}} (\bibinfo
  {year} {2023}{\natexlab{b}})\BibitemShut {NoStop}%
\bibitem [{\citenamefont {Schäfer}\ \emph {et~al.}(2023)\citenamefont
  {Schäfer}, \citenamefont {Kambs}, \citenamefont {Herrmann}, \citenamefont
  {Bauer},\ and\ \citenamefont {Becher}}]{schafer_two-stage_2023}%
  \BibitemOpen
  \bibfield  {author} {\bibinfo {author} {\bibfnamefont {M.}~\bibnamefont
  {Schäfer}}, \bibinfo {author} {\bibfnamefont {B.}~\bibnamefont {Kambs}},
  \bibinfo {author} {\bibfnamefont {D.}~\bibnamefont {Herrmann}}, \bibinfo
  {author} {\bibfnamefont {T.}~\bibnamefont {Bauer}},\ and\ \bibinfo {author}
  {\bibfnamefont {C.}~\bibnamefont {Becher}},\ }\href
  {https://doi.org/10.48550/arXiv.2307.11389} {\bibinfo {title} {Two-stage, low
  noise quantum frequency conversion of single photons from silicon-vacancy
  centers in diamond to the telecom {C}-band}} (\bibinfo {year}
  {2023})\BibitemShut {NoStop}%
\bibitem [{\citenamefont {Bersin}\ \emph {et~al.}(2023)\citenamefont {Bersin},
  \citenamefont {Sutula}, \citenamefont {Huan}, \citenamefont {Suleymanzade},
  \citenamefont {Assumpcao}, \citenamefont {Wei}, \citenamefont {Stas},
  \citenamefont {Knaut}, \citenamefont {Knall}, \citenamefont {Langrock},
  \citenamefont {Sinclair}, \citenamefont {Murphy}, \citenamefont {Riedinger},
  \citenamefont {Yeh}, \citenamefont {Xin}, \citenamefont {Bandyopadhyay},
  \citenamefont {Sukachev}, \citenamefont {Machielse}, \citenamefont
  {Levonian}, \citenamefont {Bhaskar}, \citenamefont {Hamilton}, \citenamefont
  {Park}, \citenamefont {Lončar}, \citenamefont {Fejer}, \citenamefont
  {Dixon}, \citenamefont {Englund},\ and\ \citenamefont
  {Lukin}}]{bersin_telecom_2023}%
  \BibitemOpen
  \bibfield  {author} {\bibinfo {author} {\bibfnamefont {E.}~\bibnamefont
  {Bersin}}, \bibinfo {author} {\bibfnamefont {M.}~\bibnamefont {Sutula}},
  \bibinfo {author} {\bibfnamefont {Y.~Q.}\ \bibnamefont {Huan}}, \bibinfo
  {author} {\bibfnamefont {A.}~\bibnamefont {Suleymanzade}}, \bibinfo {author}
  {\bibfnamefont {D.~R.}\ \bibnamefont {Assumpcao}}, \bibinfo {author}
  {\bibfnamefont {Y.-C.}\ \bibnamefont {Wei}}, \bibinfo {author} {\bibfnamefont
  {P.-J.}\ \bibnamefont {Stas}}, \bibinfo {author} {\bibfnamefont {C.~M.}\
  \bibnamefont {Knaut}}, \bibinfo {author} {\bibfnamefont {E.~N.}\ \bibnamefont
  {Knall}}, \bibinfo {author} {\bibfnamefont {C.}~\bibnamefont {Langrock}},
  \bibinfo {author} {\bibfnamefont {N.}~\bibnamefont {Sinclair}}, \bibinfo
  {author} {\bibfnamefont {R.}~\bibnamefont {Murphy}}, \bibinfo {author}
  {\bibfnamefont {R.}~\bibnamefont {Riedinger}}, \bibinfo {author}
  {\bibfnamefont {M.}~\bibnamefont {Yeh}}, \bibinfo {author} {\bibfnamefont
  {C.~J.}\ \bibnamefont {Xin}}, \bibinfo {author} {\bibfnamefont
  {S.}~\bibnamefont {Bandyopadhyay}}, \bibinfo {author} {\bibfnamefont {D.~D.}\
  \bibnamefont {Sukachev}}, \bibinfo {author} {\bibfnamefont {B.}~\bibnamefont
  {Machielse}}, \bibinfo {author} {\bibfnamefont {D.~S.}\ \bibnamefont
  {Levonian}}, \bibinfo {author} {\bibfnamefont {M.~K.}\ \bibnamefont
  {Bhaskar}}, \bibinfo {author} {\bibfnamefont {S.}~\bibnamefont {Hamilton}},
  \bibinfo {author} {\bibfnamefont {H.}~\bibnamefont {Park}}, \bibinfo {author}
  {\bibfnamefont {M.}~\bibnamefont {Lončar}}, \bibinfo {author} {\bibfnamefont
  {M.~M.}\ \bibnamefont {Fejer}}, \bibinfo {author} {\bibfnamefont {P.~B.}\
  \bibnamefont {Dixon}}, \bibinfo {author} {\bibfnamefont {D.~R.}\ \bibnamefont
  {Englund}},\ and\ \bibinfo {author} {\bibfnamefont {M.~D.}\ \bibnamefont
  {Lukin}},\ }\href {http://arxiv.org/abs/2307.08619} {\bibinfo {title}
  {Telecom networking with a diamond quantum memory}} (\bibinfo {year}
  {2023})\BibitemShut {NoStop}%
\bibitem [{\citenamefont {Padrón-Brito}\ \emph {et~al.}(2021)\citenamefont
  {Padrón-Brito}, \citenamefont {Lowinski}, \citenamefont {Farrera},
  \citenamefont {Theophilo},\ and\ \citenamefont
  {de~Riedmatten}}]{padron-brito_probing_2021}%
  \BibitemOpen
  \bibfield  {author} {\bibinfo {author} {\bibfnamefont {A.}~\bibnamefont
  {Padrón-Brito}}, \bibinfo {author} {\bibfnamefont {J.}~\bibnamefont
  {Lowinski}}, \bibinfo {author} {\bibfnamefont {P.}~\bibnamefont {Farrera}},
  \bibinfo {author} {\bibfnamefont {K.}~\bibnamefont {Theophilo}},\ and\
  \bibinfo {author} {\bibfnamefont {H.}~\bibnamefont {de~Riedmatten}},\ }\href
  {https://doi.org/10.1103/PhysRevResearch.3.033287} {\bibfield  {journal}
  {\bibinfo  {journal} {Physical Review Research}\ }\textbf {\bibinfo {volume}
  {3}},\ \bibinfo {pages} {033287} (\bibinfo {year} {2021})}\BibitemShut
  {NoStop}%
\bibitem [{\citenamefont {Massar}\ and\ \citenamefont
  {Popescu}(1995)}]{massar_optimal_1995}%
  \BibitemOpen
  \bibfield  {author} {\bibinfo {author} {\bibfnamefont {S.}~\bibnamefont
  {Massar}}\ and\ \bibinfo {author} {\bibfnamefont {S.}~\bibnamefont
  {Popescu}},\ }\href {https://doi.org/10.1103/PhysRevLett.74.1259} {\bibfield
  {journal} {\bibinfo  {journal} {Physical Review Letters}\ }\textbf {\bibinfo
  {volume} {74}},\ \bibinfo {pages} {1259} (\bibinfo {year}
  {1995})}\BibitemShut {NoStop}%
\bibitem [{\citenamefont {Specht}\ \emph {et~al.}(2011)\citenamefont {Specht},
  \citenamefont {Nölleke}, \citenamefont {Reiserer}, \citenamefont {Uphoff},
  \citenamefont {Figueroa}, \citenamefont {Ritter},\ and\ \citenamefont
  {Rempe}}]{specht_single-atom_2011}%
  \BibitemOpen
  \bibfield  {author} {\bibinfo {author} {\bibfnamefont {H.~P.}\ \bibnamefont
  {Specht}}, \bibinfo {author} {\bibfnamefont {C.}~\bibnamefont {Nölleke}},
  \bibinfo {author} {\bibfnamefont {A.}~\bibnamefont {Reiserer}}, \bibinfo
  {author} {\bibfnamefont {M.}~\bibnamefont {Uphoff}}, \bibinfo {author}
  {\bibfnamefont {E.}~\bibnamefont {Figueroa}}, \bibinfo {author}
  {\bibfnamefont {S.}~\bibnamefont {Ritter}},\ and\ \bibinfo {author}
  {\bibfnamefont {G.}~\bibnamefont {Rempe}},\ }\href
  {https://doi.org/10.1038/nature09997} {\bibfield  {journal} {\bibinfo
  {journal} {Nature}\ }\textbf {\bibinfo {volume} {473}},\ \bibinfo {pages}
  {190} (\bibinfo {year} {2011})}\BibitemShut {NoStop}%
\bibitem [{\citenamefont {Raa}\ \emph {et~al.}(2023)\citenamefont {Raa},
  \citenamefont {Ervasti}, \citenamefont {Botma}, \citenamefont {Visser},
  \citenamefont {Budhrani}, \citenamefont {van Rantwijk}, \citenamefont
  {Cadot}, \citenamefont {Vermeltfoort}, \citenamefont {Pompili}, \citenamefont
  {Stolk}, \citenamefont {Weaver}, \citenamefont {van~der Enden}, \citenamefont
  {de~Leeuw~Duarte}, \citenamefont {Teng}, \citenamefont {van Zwieten},\ and\
  \citenamefont {Grooteman}}]{raa_qmi_2023}%
  \BibitemOpen
  \bibfield  {author} {\bibinfo {author} {\bibfnamefont {I.~T.}\ \bibnamefont
  {Raa}}, \bibinfo {author} {\bibfnamefont {H.~K.}\ \bibnamefont {Ervasti}},
  \bibinfo {author} {\bibfnamefont {P.~J.}\ \bibnamefont {Botma}}, \bibinfo
  {author} {\bibfnamefont {L.~C.}\ \bibnamefont {Visser}}, \bibinfo {author}
  {\bibfnamefont {R.}~\bibnamefont {Budhrani}}, \bibinfo {author}
  {\bibfnamefont {J.~F.}\ \bibnamefont {van Rantwijk}}, \bibinfo {author}
  {\bibfnamefont {S.~P.}\ \bibnamefont {Cadot}}, \bibinfo {author}
  {\bibfnamefont {J.}~\bibnamefont {Vermeltfoort}}, \bibinfo {author}
  {\bibfnamefont {M.}~\bibnamefont {Pompili}}, \bibinfo {author} {\bibfnamefont
  {A.~J.}\ \bibnamefont {Stolk}}, \bibinfo {author} {\bibfnamefont {M.~J.}\
  \bibnamefont {Weaver}}, \bibinfo {author} {\bibfnamefont {K.~L.}\
  \bibnamefont {van~der Enden}}, \bibinfo {author} {\bibfnamefont
  {D.}~\bibnamefont {de~Leeuw~Duarte}}, \bibinfo {author} {\bibfnamefont
  {M.}~\bibnamefont {Teng}}, \bibinfo {author} {\bibfnamefont {J.}~\bibnamefont
  {van Zwieten}},\ and\ \bibinfo {author} {\bibfnamefont {F.}~\bibnamefont
  {Grooteman}},\ }\href
  {https://doi.org/10.4121/6D39C6DB-2F50-4A49-AD60-5BB08F40CB52} {\bibinfo
  {title} {{QMI} - {Quantum} {Measurement} {Infrastructure}, a {Python} 3
  framework for controlling laboratory equipment}} (\bibinfo {year}
  {2023})\BibitemShut {NoStop}%
\end{thebibliography}%

\title{Supplementary Information for \\ ``Qubit teleportation between a memory-compatible photonic time-bin qubit and a solid-state quantum network node"}
\maketitle

\onecolumngrid

\section{Frequency Locking}
\begin{figure}[h]
    \centering
    \includegraphics[width = \textwidth]{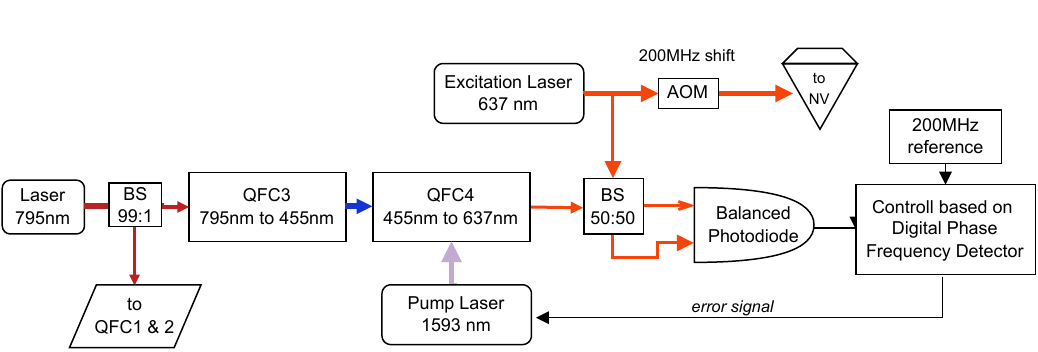}
    \caption{Outline of the frequency locking scheme.}
    \label{fig:freq-lock}
\end{figure}

To ensure indistinguishability in frequency between the converted weak coherent states and the zero phonon line emission of the NV we apply an active frequency locking scheme outlined in Fig. \ref{fig:freq-lock}. We use two additional frequency converters (QFC3 and QFC4) which convert a continuous wave tap-off from the 795nm laser to light at 637nm. We interfere this light on a balanced beamsplitter with light from the laser that excites the NV center. This excitation light passes through an AOM before reaching the NV center, which shifts the frequency by 200 MHz. Thus, we stabilize the frequency of the converted light to a fixed frequency offset of 200 MHz. We detect the light in the two output ports of the beamsplitter (BS) using a balanced photodiode, the output of which feeds, together with a 200 MHz reference, into a custom control box based on a HMC3716 Digital Phase Frequency Detector. This control box generates an error signal which we use to adapt the frequency of our telecom pump laser.

\section{Model of Expected TPQI Visibility}
In this section we will give details on our model to predict the Two Photon Quantum Interference (TPQI) Visibility. We base this model on a similar one presented in \cite{padron-brito_probing_2021}. We will derive the expected TPQI visibility as a function of $p_{NV}$, the NV emission probability and $|\alpha|^2$  the mean photon number of the weak coherent state.

We start by defining the visibility as 

\begin{equation}    
V=1-\dfrac{p_{ind}}{p_{dist}}
\end{equation}

with $p_{ind}$ ($p_{dist}$) the probability of a coincidence detection if the photons are indistinguishable (distinguishable).
We will assume the probability of 3-photon events to be negligible and thus $p_{ind}$ consists of four contributions
\begin{equation}
     p_{ind}= p_{2nv} + p_{2wcs} + p_{nv,wcs} + p_{bg},
\end{equation}
where $p_{2nv}$ is the probability of detecting 2 NV photons, $p_{2wcs}$ is the probability of detecting 2 photons from the weak coherent state, $p_{nv,wcs}$ is the probability of detecting one NV photon and one weak coherent state photon and $p_{bg}$ is the probability of detecting a coincidence where one click is originated from the background noise.
Here, the probability of detecting 2 photons in the NV detection window is given by
\begin{equation}
    p_{2nv} = \dfrac{1}{4}p^2_{NV} g^{(2)}
\end{equation}
The autocorrelation coefficient $g^{(2)}$ of the NV can be determined separately during the distinguishable sequence of the TPQI experiment by calculating the ratio of a coincidence event $p_{coinc}$ between the two detectors $D1$ and $D2$ and the individual probabilities of a detector click $p_{D1(2)}= 0.5 p_{NV}$, 
$g^{(2)} = \frac{p_{coinc}}{p_{D1}p_{D2}}$.

Secondly, the probability of detecting two photons originating from the weak coherent state is given by
$p_{2wcs} = \frac{1}{4} |\alpha|^4$. Furthermore, the probability of a coincidence originating from one photon entering on each side of the beam splitter depends on the indistinguishability $\eta$ of the photons

\begin{equation}
   p_{nv,wcs} = (1-\eta)\dfrac{p_{NV} |\alpha|^2}{2}
\end{equation}
and finally the probability of a coincidence where a noise count is involved is given by 
 \begin{equation}
 p_{bg} = p_{noise}(|\alpha|^2+p_{NV}+ p_{noise}) 
 \end{equation}
 with $p_{noise}$, the probability of a single background (noise) click in one detector per time window. 
Thus, $p_{ind}$ becomes 
\begin{equation}
p_{ind}=\dfrac{1}{4}p^2_{NV} g^{(2)} + \dfrac{1}{4}|\alpha|^4+(1-\eta)\dfrac{p_{NV} |\alpha|^2}{2} + p_{noise}(|\alpha|^2+p_{NV} +  p_{noise})
\end{equation}

For perfectly  distinguishable photons ($\eta =0$) we obtain

\begin{equation}
p_{dist}=\dfrac{1}{4}p^2_{NV} g^{(2)} + \dfrac{1}{4}|\alpha|^4+\dfrac{p_{NV} |\alpha|^2}{2} + p_{noise}(|\alpha|^2+p_{NV} +  p_{noise})
\end{equation}

This then leads to 

\begin{equation}
V=1-\dfrac{p_{ind}}{p_{dist}} = 1 - \dfrac{\dfrac{1}{4}p^2_{NV} g^{(2)} + \dfrac{1}{4}|\alpha|^4+(1-\eta)\dfrac{p_{NV} |\alpha|^2}{2} + p_{noise}(|\alpha|^2+p_{NV} +  p_{noise})} {\dfrac{1}{4}p^2_{NV} g^{(2)} + \dfrac{1}{4}|\alpha|^4+\dfrac{p_{NV} |\alpha|^2}{2} + p_{noise}(|\alpha|^2+p_{NV} +  p_{noise})}
\end{equation}

or
\begin{equation}
V= \dfrac{\eta p_{NV} |\alpha|^2}{\dfrac{1}{2}p_{NV}^2 g^{(2)} + \dfrac{1}{2}| \alpha|^4 + p_{NV} |\alpha|^2 + 2p_{noise}(|\alpha|^2+p_{NV} +  p_{noise})}
\end{equation}

Finally, we can  re-write the Visibility as a function of the ratio $x$ between $|\alpha|^2$ and $p_{NV}$, with $x = |\alpha|^2/p_{NV}$ as follows 

\begin{equation}V= \dfrac{\eta x}{\dfrac{1}{2}g^{(2)} + \dfrac{1}{2}x^2 + x + \dfrac{2 p_{noise} (1+x)}{p_{NV}} + \dfrac{2 p_{noise}^2}{p_{NV}^2}}\end{equation}

which we use to fit the data shown in Fig. 3 of the main text. 
\begin{figure}[ht!]
    \centering
    \includegraphics[width=0.8\linewidth]{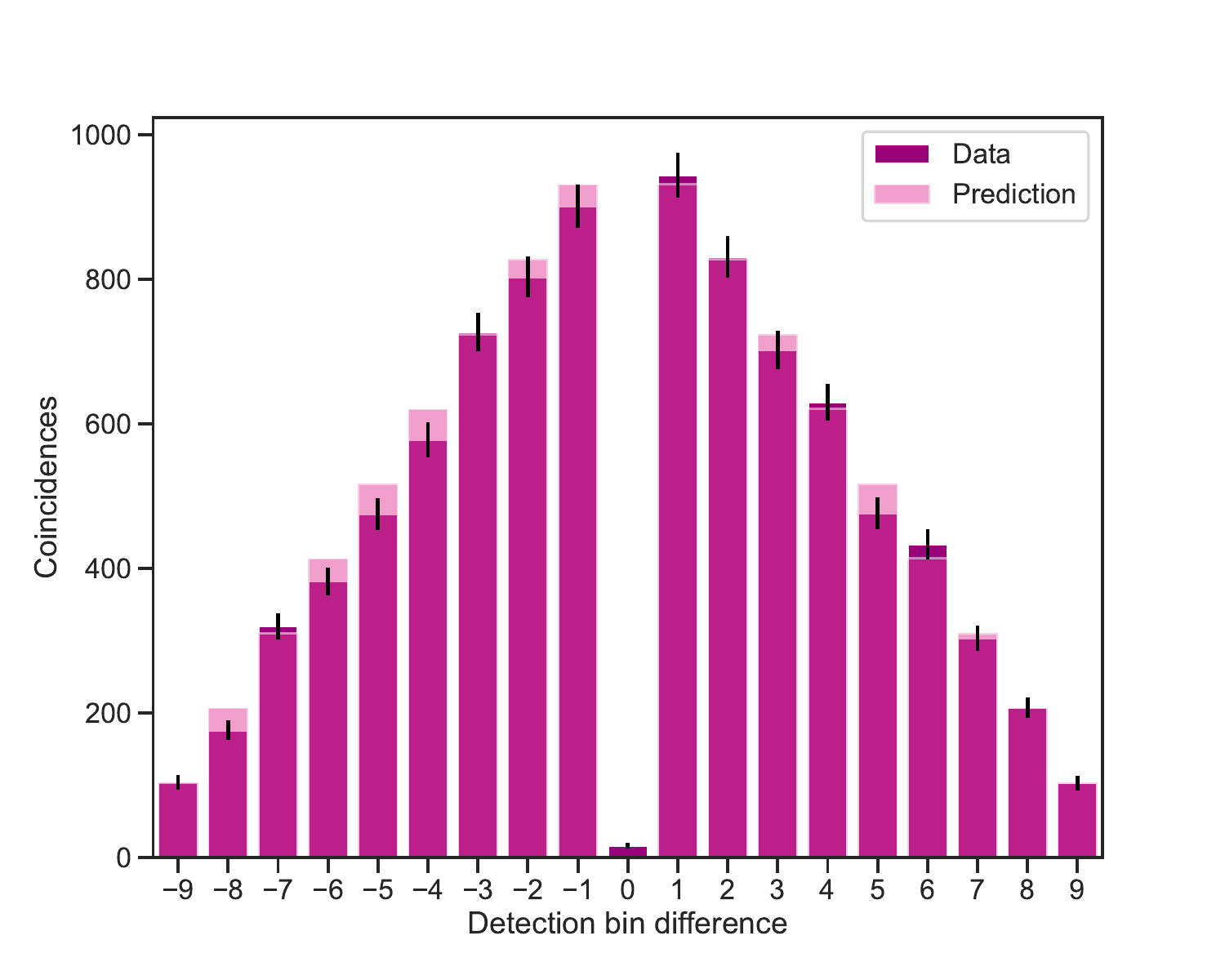}
    \caption{Autocorrelation function of the NV center. The magenta bins represent the data collected during the TPQI experiment for all the measured ratios. The pink bins, instead, represent the expected coincidences when assuming a perfect single-photon source. The resulting g$^{(2)}$ value corrected for noise is 0.011$\pm$0.004.}
    \label{fig:g2}
\end{figure}

\section{Model of Expected Teleportation Fidelity}


In this section we will derive the model we used to predict the Fidelity of the teleported state. We start by defining the input states and measurements and then continue to discuss the different emission patterns that can lead to a valid heralding event and how they affect the final fidelity we can expect to observe. 

The state to be teleported will be denoted as $|\Psi_{A}\rangle$ and is prepared in either the early time bin $|E\rangle$, the late time bin $|L\rangle$ or an equal superposition of the two. The set of states prepared for teleportation are the two states on the poles of the Bloch-sphere
\begin{align}
&|+Z\rangle = |E\rangle \\
&|-Z\rangle = |L\rangle \\
\end{align}
as well as the four equatorial states 
\begin{align}
&|\pm X\rangle = \frac {1}{\sqrt{2}}(|E\rangle \pm |L\rangle) \\
&|\pm Y\rangle = \frac {1}{\sqrt{2}}(|E\rangle \pm i |L\rangle) \\
\end{align}. 
Due to imperfections in the preparation, the polar states cannot be prepared perfectly. Still, there is a probability of leakage light emission in the orthogonal time bin which we will consider in our model. 
The electron spin qubit of the NV center and the emitted photon are prepared in the joint state 
\begin{equation}|\Phi\rangle_{B} = \dfrac{1}{\sqrt{2}} (|1\rangle| E\rangle + |0\rangle |L\rangle)\end{equation}
where $|0\rangle$ ($|1\rangle$) denotes the bright (dark) state of the spin qubit (see main text).

The Bell-state measurement, required for teleportation consists of interfering the two photonic states on a balanced beam splitter and post-selecting events where a detection event happened both in the early and late time-bin (either in the same or in different detectors). This corresponds to projection onto the states:
\begin{equation}|\Psi^{\pm} \rangle = \dfrac{1}{\sqrt{2}} (|E\rangle | L\rangle \pm |L\rangle |E\rangle)\end{equation}
where the sign is determined by the detection pattern. 
As neither of the photon sources emits perfect single-photon states, we have to consider several distinct emission cases that can lead to an accepted heralding signal and how these affect the resulting quantum state. Due to the low emission probability, we neglect all terms in which more than two photons are emitted from any side as well as the case where both sides emit two photons. 
We will start by defining the different probabilities of occurrence for different photon numbers. For the weak coherent state we have: 
\begin{align}
P^{w}_0 &= e^{-\mu} \\
P^{w}_1 &= e^{-\mu} \mu \\
P^{w}_2 &= e^{-\mu} \dfrac{\mu ^ 2}{2}\\
\end{align}
with $P_i^w$the probability of emission of $i$ photons from a weak coherent state with mean photon number $\mu$.
 
The probability of collecting $i$ photons from the NV-center is given by $P^{NV}_i$ 
\begin{align}
P^{NV}_0 &= (1-p_{NV})\\
P^{NV}_1 &= p_{NV} (1-p_{de})\\
P^{NV}_2 &= p_{NV} p_{de}\\
\end{align}
where $p_{de}$ denotes the double excitation probability of the NV-center.
We now write out these individual contributions as non-normalized density matrices $\rho_{ij}$ with their respective probability of occurrence $P_{i}^{NV} P_j^{w}$  for the different numbers of photons emitted from the NV center ($i$) and from the weak coherent state ($j$) as well as contributions in which (at least) one click was triggered by a background (or noise) count. For the case where the weak coherent state is prepared in a pole state ($\pm |Z\rangle$) we also consider the probability of emitting $k$ unwanted or leaked photons in the state orthogonal to the desired one which are denoted as $P^{w \perp}_k$.
In the following we use the simplified notation $P_{ijk}$ for $P^{NV}_i P^{w}_j P^{w\perp}_k  $ or $P_{ij}$ for $P^{NV}_i P^{w}_j$.
The probability of one or two background or noise photons contributing to a valid trigger event is, for the pole states, given by 
\begin{align}
&P_{bg}^{pole}  = 2p_{noise}(2p_{noise}P_{000} + P_{010} + P_{100}+ P_{001})
\end{align} 
and for the equatorial states by 
\begin{align}
&P_{bg}^{eq}  = 2p_{noise}(2p_{noise}P_{00}  + P_{10} + P_{01})
\end{align} 
Where $p_{noise}$ is the probability of a background or noise detection per detector and time bin and we have limited the background contributions we consider to a maximum of two emitted photons. For the teleportation experiment, the measured p$_{noise}$ per detector is (5.5$\pm$0.2)e$^{-6}$.

Now we will consider the non-normalized density matrix contributions corresponding to these probabilities of occurrence. They are non-normalized as not all detection patterns are considered valid trigger events and we will post-select on these valid heralding events.  
In writing down these contributions and their probabilities of yielding a valid heralding event,   we will have to differentiate between teleporting the pole-states ($|\pm Z\rangle$) from the states in the equatorial plane of the Bloch-sphere ($|\pm X\rangle$ and $|\pm Y\rangle$), which we will mark as $\rho^{pole}$
and $\rho^{eq}$.
For the desired case of one emitted NV photon and one photon emitted from the weak coherent state we can write

\begin{align}
&\rho^{pole}_{11} = \frac{P_{110} }{2}|\Psi_A\rangle \langle \Psi_A| + \frac{P_{101} }{2}|\Psi_A^{\perp}\rangle \langle \Psi_A^{\perp}| \\
&\rho^{eq}_{11} = P_{11}(\frac{\eta}{2} |\Psi_A\rangle \langle \Psi_A|+ \frac{(1- \eta)}{2}\mathbb{I} )
\end{align}
The factor $\frac{1}{2}$ takes into account the fact that we could project on any of the four Bell-states but we can only unambiguously discern two of them and thus, only these two will lead to an accepted heralding pattern. The difference between the pole and the equatorial states is due to the fact that in the case of the pole states we profit from the classical correlations in the system, while for the equatorial states we will only obtain the desired result if the photons from the two sources interfere. 

In the case of a double emission from the weak coherent state and no NV photon, there will only be a valid trigger event for a pole state if the double emission happened in the form of one photon from the desired time bin and one in the orthogonal one, and there can as well be a valid trigger in case of an equatorial state which will lead to 

\begin{align}
&\rho^{pole}_{02} = P_{011} \mathbb{I}\\ 
&\rho^{eq}_{02} = \frac{P_{02}}{2}\mathbb{I}
\end{align}

We will omit the case in which one photon was emitted from the NV center but two from the weak coherent state due to it's low probability of occurrence for the mean photon numbers used in the experiment. 

Finally, for the case of two NV photons and one from the weak coherent state we get

\begin{align}
&\rho^{pole}_{21} = \frac{P_{210}}{4} |\Psi_A\rangle \langle \Psi_A| + \frac{P_{201}}{4}|\Psi_A^{\perp} \rangle \langle \Psi_A^{\perp}| \\
&\rho^{eq}_{21} = \frac{P_{21}}{4} \mathbb{I}
\end{align}

When teleporting pole states the final density matrix becomes

\begin{align}
\rho^{pole} = &\frac{\frac{1}{2}P_{110}  + \frac{1}{4}P_{210} }{N}  |\Psi_A\rangle \langle \Psi_A|+ \frac{\frac{1}{2}P_{101} + \frac{1}{4}P_{201} }{N}  |\Psi_A^{\perp}\rangle \langle \Psi_A^{\perp}|+ \frac{P_{011} + P_{bg}^{pole}}{N}\mathbb{I}
\end{align}
with 

\begin{align}
    N = \frac{P_{110} +P_{101}}{2}+ \frac{P_{210} + P_{201}}{4} + P_{011} + P_{bg}^{pole} 
\end{align}

Thus, we can calculate the expected fidelity to be 

\begin{align}
    F^{pole} = \dfrac{P_{110} +  \frac{1}{2}P_{210} + P_{011} +  P_{bg}^{pole} }{2N}
\end{align}

For the equatorial states we obtain 
\begin{align}
\rho^{eq} = \frac{ \eta P_{11}}{2N} |\Psi_A\rangle \langle \Psi_A|+ 
 \frac{(1- \eta)P_{11} + P_{02} + P_{21} + 2P_{bg}^{eq} }{2N}\mathbb{I}    
\end{align}

with 
 
\begin{align}
N = \frac{P_{11} + P_{02} }{2} + \frac{P_{21}}{4}  + P_{bg}^{eq}, 
\end{align}

thus, the Fidelity is given as 
\begin{align}
    F^{eq} = \dfrac{1}{2N}(  \dfrac{P_{11}(1+\eta) + P_{02} + P_{21}}{2}  + P_{bg}^{eq})
\end{align}

\section{Classical bound when teleporting with weak coherent states}

When comparing our experimentally achieved teleportation fidelity with the classical bound of $\frac{2}{3}$, this bound is derived using the optimal classical strategy when using single photons to encode qubits \cite{massar_optimal_1995}. In the case of qubits encoded in weak coherent states, however, a classical strategy might use the higher photon number contributions of that state to achieve a higher probability of success. One can calculate the maximally achievable Fidelity for a classical strategy as shown in \cite{specht_single-atom_2011}
as 
\begin{align}
F_{max}(|\alpha|^2)= \sum_{N \geq 1} F_{MP}(N)\dfrac{p(|\alpha|^2, N)}{1 - p(|\alpha|^2, 0)}
\end{align}

where $N$ is the number of photons per pulse, $p$ describes the poissonian distribution of photon numbers \begin{align}
    p(|\alpha|^2,N)= \frac{|\alpha|^{2N}}{N!}e^{-|\alpha|^2}
\end{align}
and 
\begin{align}
    F_{MP}(N)=\frac{N+1}{N+2}
\end{align}
is the maximum achievable fidelity for a state with a fixed amount of photons $N$.
For the parameters of our teleportation experiment we can now calculate the maximum achievable fidelity for a classical strategy as follows:  During the teleportation experiment we had an average NV emission probability of $p_{NV} = (4.50 \pm 0.9)e^{-4}$ and a ratio $\frac{|\alpha|^2}{p_{NV}} = 1.20 \pm 0.24$. It is noteworthy that, with respect to the TPQI experiment, a degradation of the $p_{NV}$ parameter occurred, as well as lower CR check counts were encountered. The reason for the lowered photon emission of the NV might be due to a fault in our cryostat that led to ice formation inside the sample chamber. Despite that, the setup was stable in the new configuration and the experiment was executed remotely. At the same time, the intensity modulator employed for the generation of the time-bin qubits showed higher leakage, leading to a generally increased noise probability. 
Using the upper bound of this ratio between mean photon number and $p_{NV}$ (and thus $|\alpha|^2 = 6.50e^{-4}$) the maximally achievable fidelity for a classical strategy would be $F_{max}= 0.666694$. As we can see, due to the low mean photon numbers used in our experiment, the correction is minimal but should be considered in implementations with higher mean photon numbers.
\section{Noise characterization}
In this section, we report on the characterization of the noise sources that are involved in the experiments.

\begin{center}
    \begin{figure}[htb!]
        \centering
        \includegraphics[width=0.5\columnwidth]{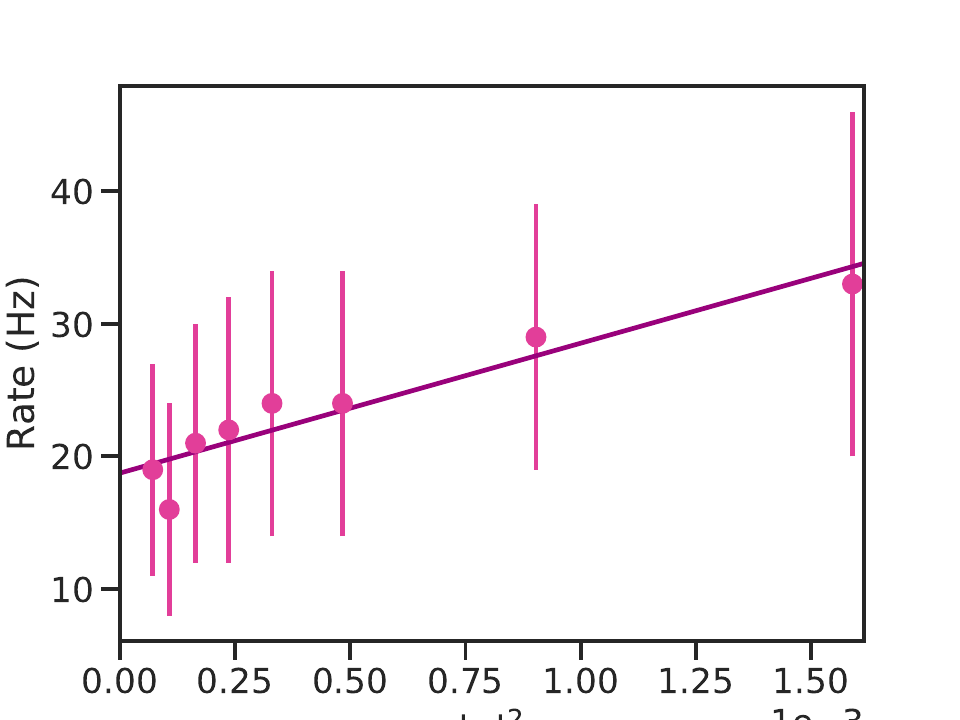}
        \caption{Noise characterization of the weak coherent state. By sweeping the voltage applied to the VOA, the mean-photon number changes. We assume a linear model for the noise count rate vs. the mean-photon number. The result of the linear fit is used in the calculation of the p$_{noise}$ term in eq. 7, thus affecting the visibility of the TPQI experiment.}
        \label{fig:wcs-noise}
    \end{figure}
\end{center}

To obtain the dark count rates of the detector, we block the ZPL collection path of the NV and the output of the QFC2. This results in a mean noise rate of (11.7$\pm$5.6)Hz.
The contribution of the pump lasers for the two-step frequency conversion is measured by keeping the ZPL path closed and blocking the 795nm input of the QFC1. The rate in this case is (12.3$\pm$5.6)Hz, showing that our conversion setup is low noise, if compared with the rate obtained when no conversion setup was involved.
To characterize the noise contribution of the weak-coherent state in the NV center window (particularly relevant in the distinguishable sequence of the TPQI experiment), we block the ZPL path of the NV center and we sweep the voltage applied to the variable optical attenuator, namely we manipulate the mean-photon number of the weak-coherent state. We play the distinguishable sequence of the TPQI experiment, collecting the counts in the time window where the NV center pulse is supposed to be. The results are illustrated in Fig. \ref{fig:wcs-noise}.\\
Lastly, we check the noise contribution coming from the NV setup. We block the output of the QFC2 setup and we open the ZPL path. We repeat the distinguishable sequence, collecting counts in the weak-coherent state window. The rate is (12.8 $\pm$ 5.4)Hz, which is comparable with the rate measured above for the detector dark counts and the pump noise. We can therefore conclude that the main source of noise comes from the preparation of the weak coherent state, particularly the combination of the intensity modulator, whose bias voltage needs to be optimized throughout the measurement, and the variable optical attenuator. However, this noise source is relevant only in the distinguishable sequence of the TPQI experiment, while in the indistinguishable one, we can consider as p$_{noise}$ the constant background noise given by the detectors and the conversion setup.
For the teleportation experiment, the noise rate increased to (275$\pm$10)Hz due to equipment degradation, as discussed above.
\section{Phase modulator characterization and stability}

To characterize the phase modulator, namely to identify $V_\pi$ and $V_{\pi/2}$, we build a Mach-Zehnder-type interferometer at 795nm. In particular, the input of the QFC1 from main text, the shaped pulses, is connected to a 50:50 in-fiber beam splitter. The two output arms of such a beam splitter are 2m long fibers, and on one of the two arms, we include the phase modulator device we want to characterize. The two arms impinge on a second in-fiber 50:50 beam splitter, whose output ports are connected to two APDs.
The voltage source for the phase modulator is one of the wave channels of the AWG, whose output has an amplitude between $\pm$5V.

\begin{figure}[h!]
    \centering
\includegraphics[width=1.0\linewidth]{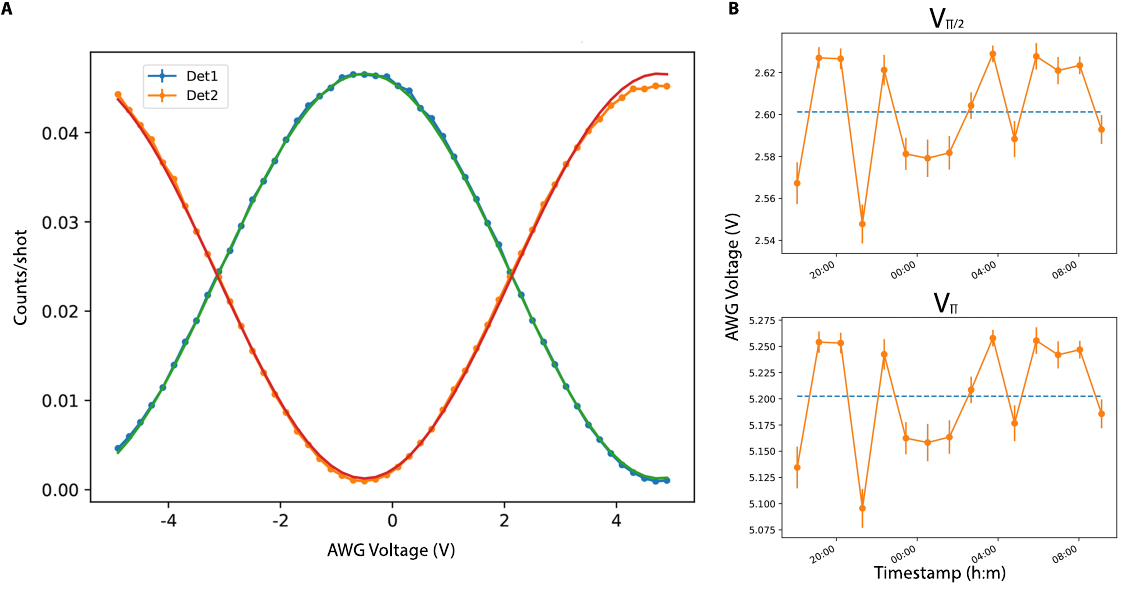}
    \caption{a) Data fit to extract the values of $V_\pi$ and $V_{\pi/2}$. The solid lines represent the fitted curve per detector. b) Stability measurement for $V_\pi$ and $V_{\pi/2}$. The dashed lines represent the average value. }
    \label{fig:phase_mod}
\end{figure}

We send a pulse to the interferometer, encountering a phase shift due to the phase modulator, and we register, through the time-tagger at the detectors, the counts per shot for the two pulses. We repeat this experiment while sweeping the voltage applied to the phase modulator, reconstructing the plot in Fig.\ref{fig:phase_mod}a for the two detectors. 

The data collected for the two detectors are jointly fit to the following curves:
$$
A_1\sin\bigg(\frac{s\cdot cps_{det_1}}{2}+o\bigg)+c_1
$$
$$
A_2\cos\bigg(\frac{s\cdot cps_{det2}}{2}+o\bigg)+c_2
$$
From the fit of the $s$ parameter, it is possible to estimate $V_\pi$ (and $V_{\pi/2}$) as $|\pi/s|$ (and $|\pi/(2s)|$). 

At this point, we repeat the measurement and the data fit more times over a time span of 15h. In Fig.\ref{fig:phase_mod}b we report the estimation of $V_{\pi/2}$ and $V_\pi$ over time, resulting in an average of (2.601$\pm$0.002)V and (5.202$\pm$0.004)V. These values are then used to make the time bin qubits in the several cardinal states. The results in Fig.\ref{fig:phase_mod}b also show the stability of these values, confirming the reproducibility. Given that $V_\pi$ exceeds the maximum amplitude that the HDAWG can provide, we use a phase modulator pulse per bin for this case.

\section{Devices and experimental monitoring}
As stated in the main text, the NV setup is ``Alice" of \cite{pompili_realization_2021, hermans_qubit_2022}. However, some devices have been replaced. In particular, in this work we employed the Zurich Instruments HDAWG as an arbitrary waveform generator, and the PicoQuant MultiHarp as timetagger. 
All the other devices remained unchanged. 

The experiments, both the TPQI and the teleportation, were running remotely, at a distance of up to 7745km between the scientists and the setup. This shows that the setup is robust over time and automated experimental monitoring is crucial. Here we report the list of automatization routines that have been implemented using the software environment in Ref.\cite{raa_qmi_2023}.

\begin{itemize}
    \item For the TPQI experiment only: auto-relocking system for the 795nm laser. During the TPQI experiment, the 795nm laser wavelength was locked to an external cavity. A homemade background program detects when the laser frequency drifts and about to go out of lock and adjusts the piezo voltage of the laser to bring the desired spectral mode back. For the teleportation experiment, the laser was locked to a wavemeter with a constant PID loop running to keep the desired wavelength. It is important to notice that the wavelength of the 795nm laser did not change between the two experiments, as the same wavemeter was monitoring the wavelength during the TPQI experiment.
\item The system is automatically calibrated over time. In particular, the calibration is targeted at the laser power, the position of the NV with respect to the objective, in such a way as to maximize the fluorescence under the excitation using green light in the Phonon-Side Band (PSB). Small drifts in the optics of the NV setup are compensated by a Python-controlled deformable mirror that is included in the Zero-Phonon Line (ZPL) path. The calibration maximizes the fluorescence in the ZPL when the green light is on. A system of automated waveplates minimizes the leakage of pulse light in the ZPL. The bias voltage of the EOM for the NV optical $\pi$-pulses and of the intensity modulator for the 795nm pulses is also optimized during the experiments to minimize leakage. 

\item A set of control scripts checks for frequency shifts of converted light and all the lasers involved with "real-time" data filtering. In particular, when the control scripts detect an anomaly in one of the frequencies monitored via the wavemeter, a flag is raised and the ongoing measurement is tagged as failed and discarded.
\item For teleportation experiment only: calibration of the ratio mean-photon number/counts per shot of the NV every 5 datasets taken. The variable optical attenuator is controlled via the Micro-Controller Unit (MCU). In this way, it is also possible to compensate for drifts in the mean-photon number as well as in the conversion setup that might cause lower efficiency during the experiment. 
\end{itemize}

\end{document}